\documentclass[nofootinbib,prd,aps]{revtex4}
\usepackage{amsmath}
\usepackage{amsfonts}
\usepackage{amssymb}
\usepackage{dsfont}
\usepackage{mathrsfs}
\usepackage{graphicx}
\usepackage{epstopdf}
\usepackage{pstricks}
\usepackage{color}
\usepackage[all]{xy}
\usepackage{url}
\usepackage{hyperref}
\hypersetup{
    colorlinks=true, 
    linktoc=page,
    linkcolor=blue,  
    urlcolor=black
}

\newcommand{\hphi}{\hat{\varphi}}
\newcommand{\hphid}{\hat{\varphi}^{\dagger}}

\newcommand{\hsigma}{\hat{\sigma}}
\newcommand{\hsigmad}{\hat{\sigma}^{\dagger}}

\newcommand{\suinter}[6]{\iota^{j_{#1}j_{#2}j_{#3}j_{#4}#5}_{#6_{#1}#6_{#2}#6_{#3}#6_{#4}}}

\newcommand{\move}{\widehat{\mathcal{M}}}

\newcommand{\tr}{\mathrm{Tr}}

\newcommand{\fv}{\left| 0 \right\rangle}

\newcommand{\ket}[1]{\left| #1 \right \rangle}

\newcommand{\extd}{\mathrm{d}}

\newcommand{\SU}{\mathrm{SU}}

\newcommand{\va}{\scriptscriptstyle}

\newcommand{\be}{\nopagebreak[3]\begin{equation}}
\newcommand{\ee}{\end{equation}}

\newcommand{\bee}{\nopagebreak[3]\begin{equation*}}
\newcommand{\eee}{\end{equation*}}

\newcommand{\ba}{\nopagebreak[3]\begin{eqnarray}}
\newcommand{\ea}{\end{eqnarray}}

\newcommand{\baa}{\nopagebreak[3]\begin{eqnarray*}}
\newcommand{\eaa}{\end{eqnarray*}}

\newcommand{\la}{\label}
\newcommand{\n}{\nonumber}

\newcommand{\C}{\mathbb{C}}
\newcommand{\N}{\mathbb{N}}

\newcommand{\iu}{\mathrm{i}}

\newcommand{\ie}{\emph{i.e.~}}

\newcommand{\bdyp}{$+$}
\newcommand{\bdym}{$-$}

\newcommand{\aih}{\mathcal{A}_{\va IH}}

\begin{document}

\title{Black Holes as Quantum Gravity Condensates}
\author{Daniele Oriti$^1$}
\email{daniele.oriti@aei.mpg.de}

\author{Daniele Pranzetti$^2$}
\email{dpranzetti@perimeterinstitute.ca}

\author{Lorenzo Sindoni$^3$}
\email{lorenzo.sindoni@invenialabs.co.uk}

\affiliation{$^1$Max Planck Institute for Gravitational Physics (AEI), 
Am M\"uhlenberg 1, D-14476 Golm, Germany}

\affiliation{$^2$Perimeter Institute for Theoretical Physics, 31 Caroline St N, Waterloo, ON N2L 2Y5, Canada}

\affiliation{$^3$ InveniaLabs, Parkside Place, Parkside, CB1 1HQ Cambridge, United Kingdom}

\begin{abstract}
We model spherically symmetric black holes within the group field theory formalism for quantum gravity via generalised condensate states, involving sums over arbitrarily refined graphs (dual to 3d triangulations). The construction relies heavily on both the combinatorial tools of random tensor models and the quantum geometric data of loop quantum gravity, both part of the group field theory formalism. Armed with the detailed microscopic structure, we compute the entropy associated with the black hole horizon, which turns out to be equivalently the Boltzmann entropy of its microscopic degrees of freedom and the entanglement entropy between the inside and outside regions. We recover the area law under very general conditions, as well as the Bekenstein-Hawking formula. The result is also shown to be generically independent of any specific value of the Immirzi parameter.
\end{abstract}

\maketitle

\section{Introduction}
\label{sec:Intro}
The physics of black holes is a crucial part of the current frontier of theoretical physics. Beside their importance for theoretical (and observational) astrophysics \cite{Bambi:2016lkv, Rezzolla:2016jxw}, they offer a precious laboratory for testing the foundations of our theories of fundamental interactions and of all other physical systems, and their mutual (in)compatibility: General Relativity, Quantum Mechanics (and its offspring, Quantum Information), and Statistical Mechanics. On the one hand, black hole thermodynamics 
\cite{Bekenstein:1972tm,Bardeen:1973gs,Bekenstein:1973ur,Hawking:1974rv,
Hawking:1974sw} 
(see also \cite{Carlip:2014pma} for a recent review and references covering the various developments)
remains to date a surprising set of (theoretical) facts, still in search for a microscopic (statistical) understanding. 
It constantly challenges our assumed foundations of physical theories, at least when framed within semi-classical physics: locality, conservation of information, the equivalence principle, Lorentz invariance, quantum monogamy, causality, to name only a few of the principles that have been suggested to be somehow in conflict with it. On the other hand, together with
the cosmology of the very early universe, black holes (and their thermodynamics) are 
especially important for quantum gravity research, representing the main concrete challenge 
and the first testing ground at the non-perturbative level. Quantum gravity is supposed to 
complete the very definition of what black holes are, by providing the new physics replacing 
their central curvature singularity, and to identify the microscopic degrees of freedom whose 
statistical mechanics is ultimately responsible for their macroscopic thermodynamical 
properties. 

In this paper, we tackle the issue of defining quantum states representing (spherically 
symmetric) black holes, and in particular their horizon degrees of freedom, within a full 
quantum gravity formalism, and to compute their macroscopic entropy from first principles, 
within the same formalism. The quantum gravity framework we use is group field theory 
\cite{Oriti:2013aqa,Oriti:2014uga}, 
closely related to random tensor models \cite{Gurau:2011xp} and to loop quantum gravity \cite{Rovelli:2004tv,Thiemann:2007zz}. To clarify 
what is the new contribution we give here to this topic, it is in fact useful to relate and compare 
our work to the one done so far within loop quantum gravity, forming by now an extensive 
literature.

In the canonical loop quantum gravity (LQG) approach to black hole entropy calculation, the main strategy to the computation of black hole entropy has been based on classical symmetry reduction. The starting point for the modelling the black hole horizon is the local definition provided by the notion of Isolated Horizon (IH) \cite{Ashtekar:1998sp, Ashtekar:1999yj} (see \cite{DiazPolo:2011np, Perez:2017cmj} for reviews), which enters as a specific set of boundary conditions. Their implication is that a spherically symmetric isolated horizon appears at any $r_0$ for which the following classical geometrical relation holds \cite{Ashtekar:1999wa, Engle:2010kt}:
\ba\la{IH-cond}
\mathcal{C}_{\va IH}\equiv{F}^i(A)+\frac{\pi }{\aih}(1-\gamma^2) {\Sigma}^i\la{FSigma}=0\,.
\ea
In the expression above, $\aih$ is the area of the isolated horizon, $F^i(A)$ the field strength of the Ashtekar--Barbero connection 
$A^i_a=\Gamma^i_a+\gamma K^i_a$ on the IH and $\Sigma^i\equiv \epsilon^i\,_{jk}\Sigma^{jk}$ the 2-form densitized triad pulled-back on the horizon. Spherically symmetric geometries constrained by such boundary conditions are then taken to \emph{define} the classical degrees of freedom of the black hole system, which is then quantized. 

In the quantum theory, the  IH condition \eqref{FSigma} induces a relation between the flux associated to a link coming from the bulk and ending on the horizon and the holonomy on the horizon around the given link. This is imposed as an operatorial constraint equation on the tensor product of the bulk Hilbert space, quantised through standard LQG techniques, and the boundary Hilbert space, constructed by relying on Chern-Simons theory formalism, which also defines the boundary dynamics. As a result, fluctuations of geometrical operators coming from the bulk get coupled to those living on the boundary, through identification of quantum numbers of the two Hilbert spaces, and Chern-Simons curvature excitations (\lq punctures\rq) get identified with quanta of space degrees of freedom. The microcanonical counting of these degrees of freedom \`a la Boltzmann in the semi-classical limit of large IH area yields a leading term for the entropy linear in  $\aih$ together with a subleading logarithmic term \cite{Ashtekar:2000eq, Meissner:2004ju, Domagala:2004jt, Ghosh:2004wq, Corichi:2006wn, Corichi:2006bs, Agullo:2008yv, Kaul:2000kf, G.:2008mj, Engle:2011vf}. Extension of the $\SU(2)$-invariant formulation of isolated horizons to the distorted and the rotating cases has been achieved in
\cite{Perez:2010pq, Frodden:2012en}.

The realisation that, in this classically reduced context, the relevant degrees of freedom are associated to punctures on the horizon (described by a simple Chern-Simons theory) combined with the general key result obtained in the full LQG theory that area operator has discrete spectrum, with spin networks as eigenstates and eigenvalues carried by the links of their supporting graphs, has motivated another well-explored strategy  (in fact, the first one to be followed \cite{Smolin:1995vq, Krasnov:1996wc, Rovelli:1996dv}). This was the construction of several toy models within LQG, \ie simple spin network states incorporating enough features of the mentioned description of quantized isolated horizons to have a chance to capture interesting black hole physics. Most of such toy models consist of spin network states based on a single fixed graph, interpreted as having a number of nodes inside a black hole horizon (encoding the bulk degrees of freedom), but often limited to a single intertwiner state, and a number of links crossing it, providing for the horizon degrees of freedom.

Both the mentioned strategies have produced very important and interesting results, and will certainly contribute to the complete understanding of the physics of quantum black holes in quantum gravity. The reliability of the results obtained within a symmetry reduced treatment is notoriously questionable, however, and the limitations of both strategies are apparent, motivating the search for a more complete description of quantum black holes within the full quantum gravity formalism.

Moreover, in all the LQG-based constructions so far, the numerical value of the Barbero--Immirzi  parameter needs to be fixed to recover exactly the coefficient $1/4$ in the Bekenstein--Hawking entropy area formula.
The long-standing issue, extensively debated in the literature \cite{Jacobson:2007uj, Ghosh:2011fc, Frodden:2012dq, Ghosh:2013iwa, Pranzetti:2013lma, Bodendorfer:2013hla, Ghosh:2014rra, Achour:2014eqa, BenAchour:2016mnn},
is whether this necessity might be signalling an incompleteness in the identification of the microscopic degrees of freedom counted in the LQG entropy  calculation, or some other limitation in the usual constructions. 
On the basis of the semi-classical nature of the Bekenstein--Hawking formula and the expectation that the Barbero--Immirzi  parameter should play no important role in the classical description of gravity,
a natural option to remove this undesired feature of the entropy calculation is that other (non-internal and so far neglected) degrees of freedom should be taken into account. 
A recent proposal within the LQG framework, discussed in \cite{Freidel:2016bxd}, uses new boundary degrees of freedom representing information channels between gravitational subsystems. 
However, the implications of the discovery of these new degrees of freedom for the black hole entropy calculation have not been investigated in detail yet.

In this paper, we will concentrate on an alternative proposal for the fundamental boundary (and bulk) degrees of freedom, working within the full quantum gravity formalism and using a construction that gives us access to a continuum description of a black hole quantum geometry. 
More precisely, we want to use the construction \cite{Oriti:2015qva} of GFT condensate states, generalising those used in a cosmological context \cite{Gielen:2013kla, Gielen:2013naa, Oriti:2016qtz, Gielen:2016dss}, and representing continuum spherically symmetric geometries within the group field theory (GFT) formalism.
Leveraging on the structure of the Fock representation of GFTs, it is possible to define a quantum black hole horizon in the full, non-truncated, theory and then compute its entropy.
A short report of some of these results appeared in \cite{Oriti:2015rwa}. Here we provide a much more detailed presentation of those calculations, as well as a more general entropy counting
which extends to a wider and possibly more physically relevant class of generalized condensates for a spherically symmetric black hole. 

We will describe in some detail the construction of the class of states that our analysis relies on, and highlight their convenient features in the following; we note upfront, however, the main limitation of our construction. Implementing the dynamics for generic GFT condensates is not as easy as in the case of those used
for cosmological applications \cite{Gielen:2013kla, Gielen:2013naa, Oriti:2016qtz, Gielen:2016dss}.
Therefore we will treat these states as kinematical trial states, under the hypothesis that they can represent some reasonable approximation
of realistic states, \ie solutions of the full quantum dynamics, at least in some regime.

This remains an hypothesis, but it is supported by three considerations. First, these
states naturally include some form of homogeneity (specifically, wavefunction homogeneity, to be clarified in the following) which already
restricts the possible shape of the states in combination with the combinatorial restrictions that allow to assign them a clear topological interpretation (in particular, as encoding spherical symmetry). This sort of restrictions is expected to apply also to an exact state (\ie solving the equations of motion of the theory) with spherical symmetry, or some 
slightly less local variant (e.g. involving vertices along tangential loops), resulting from 
constraints on curvature observables. Second, very importantly (and for the first time, to the best of our knowledge), the quantum states we use already include a sum over 
a family of triangulations, which is an inevitable consequence of the interacting nature of the GFT
equations of motion. It is not obvious that the superposition of different graphs 
realised in physical states is of the same melonic nature that we are using, even if the dominance of melonic diagrams is a recurrent property of GFT (and random tensor) models \cite{Baratin:2013rja}; however, the superposition that we are using  explores a whole family of triangulations obtained
by refinement, which means that, while differing in some crucial aspects, we might be
still close to the exact state (indeed, for general wavefunctions, states associated to
different graphs are \emph{not} orthogonal \cite{Oriti:2014uga}). While the calculation of the overlap
between our condensate states and an exact spherically symmetric state (assuming this notion makes sense in the full theory) is nontrivial,
we can expect that this overlap is nonzero\footnote{Its precise value would be a quantitative measure of the goodness of the approximation used.}.
Third, these particular states implement a non-obvious macroscopic property, \ie holography, in a form compatible with a non-perturbative full quantum gravity regime. This means that
they contain some of the structural properties necessary to match the dynamics of the classical theory.

Still, it remains true that our analysis stays at the kinematical level (as in all the previous LQG literature on the subject). We will use, however, a possible proxy for the dynamics of the theory, because we will impose a condition of maximisation of the entropy that one could expect to be satisfied by solutions of the quantum dynamics and, in particular, by black hole configurations.
The (maximal) entropy we will compute for our states will be shown to be interpretable both as a Boltzmann entropy of horizon degrees of freedom and as an entanglement entropy across the same horizon, to scale with the mean area of the black hole horizon, and to match (under additional assumptions, amounting to semi-classicality conditions) the Bekenstein--Hawking result.

We depart from the canonical LQG approach to black hole entropy in three main ways: 
$i)$ the horizon quantum state is defined including a sum over triangulations, including very refined ones, admitting in this way an interpretation in terms of a continuum geometry, whose information is encoded in a single collective variable, \ie the condensate wavefunction; crucially, this basic aspect of the construction also implies that our states encode in their very definition a coarse-graining of microscopic degrees of freedom allowing to control them via a limited number of variables---this point is also relevant for their (lack of) dynamical character: being the result of a coarse graining, we should not expect them to be exact solutions of the microscopic quantum dynamics---;
$ii)$ the interior bulk degrees of freedom are not removed (or drastically reduced) by hand through the introduction of a single intertwiner model; space-time does not end at the horizon in our construction, but both an interior and an exterior bulk are included in the quantum state;
$iii)$ our construction relies uniquely on structures and techniques proper of the GFT formalism, \ie both boundary and bulk degrees of freedom are described in a unique, consistent way, removing  
ambiguities present in the canonical LQG approach, where spin-network states are coupled to a Chern-Simons theory on the boundary\footnote{See however 
\cite{Sahlmann:2011xu, Pranzetti:2014tla} for a more uniform treatment of bulk and boundary degrees of freedom though techniques developed 
in the context of 2+1 LQG \cite{Sahlmann:2011uh, Sahlmann:2011rv, Noui:2011im, Noui:2011aa, Pranzetti:2014xva}.}. 

Most importantly, with all the limitations of our work, and the inevitable approximations and assumptions, we work from beginning to end within the full quantum gravity formalism, without any preliminary classical symmetry reduction and with realistic quantum states of the full theory.
These points of departure from the standard treatment are also at the origin of a remarkable feature of the entropy calculation we present in this paper:
by replacing  the area operator eigenstates for the quantum isolated horizon by condensate states, the result of the entropy
calculation yields the semi-classical Bekenstein--Hawking entropy area law with no explicit dependence on the value of the
Barbero--Immirzi parameter.  

\section{Shell condensate state: group element representation}
\label{sec:BH}
We start by presenting in some detail the construction of our quantum black hole states in the GFT formalism. We will do so, in this and in the next sections, in two equivalent representations of the Hilbert (Fock) space of the theory. This serves also the purpose of showing the generality of our construction, and of our results. We take a sort of engineering approach, by building up our quantum states piece by piece, starting from the fundamental building blocks provided by the GFT formalism, \ie GFT quanta corresponding to individual spin network vertices, in turn dual to fundamental 3D simplices. Specifically, we construct spherically symmetric configurations of quantum space as glueings (along a radial direction) of homogeneous spherical shells. 

As pointed out above, in this work we build upon the construction of a continuum quantum geometry representing a spherically symmetric shell performed in \cite{Oriti:2015qva}, to which we
refer for notation and basic notions\footnote{For the reader's convenience, we have reported
the basic notion in Appendix \ref{app:GFTFock}.} 
Their construction can be easily described: we start with a seed state for a given shell, containing few quanta, and then we act upon it with a series or refinement operators in order to increase the 
number of fundamental blocks (four-valent vertices), preserving the initial topology. The same wavefunction is associated to each new fundamental block, enforcing homogeneity and symmetry.

Therefore, the GFT condensate state for a given shell is formed by an infinite superposition of graphs, each with a certain number of 4-vertices connected together. A colour $t=\{B,W\}$ is associated to 
each 4-vertex and four $SU(2)$ group elements (denoted by the letter $g$ variously decorated) are assigned to the links departing from a given 4-vertex and labelled by a number $I=\{1,2,3,4\}$.
A shell is formed of three parts: an outer boundary, an inner boundary and a bulk in between. In order to keep track of these three parts of each shell, we add a colour $s=\{+,0,-\}$ to the vertex wavefunction, labelling respectively these three parts, so that we can specify the region to which a given vertex of a shell belongs.
 
Following the conventions of \cite{Oriti:2015qva} (see also Appendix \ref{app:GFTFock}), the action of the refinement operators on the initial seed state is such that each boundary of a given shell is formed by 
open radial links all with the same colour, with outer and inner boundaries having different colours. The next step is to glue shells together. In order to do this, while still  being able to distinguish 
different shells, we need to introduce a single extra label $r\in \N$, also associated to the shell wavefunction, which can be interpreted as an effective radial coordinate. Therefore, the field operator 
associated to the fundamental building block $v$ reads
\be\la{c-field}
\hsigma_{r,t^{\va v}s^{\va v}}(h^v_I) = \int \extd g^v_I\; \sigma_{r,s^{\va v}}(h^v_Ig^v_I) \,\hphi_ {t^{\va v}}(g^v_I)\,, \quad\quad
\hsigmad_{r,t^{\va v}s^{\va v}}(h^v_I) = \int \extd g^v_I\; \overline{\sigma_{r,s^{\va v}}(h^v_Ig^v_I)} \,\hphid_ {t^{\va v}}(g^v_I)\,
\ee
 satisfying the commutation relations
\ba\la{c-comm}
\left[\hsigma_{r,t^{\va v}s^{\va v}}(h^v_I), \hsigmad_{r',t^{\va w}s^{\va w}}(h^w_I) \right]  &=&
 \delta_{r,r'}\delta_{t^{\va v}, t^{\va w}}\delta_{s^{\va v}, s^{\va w}}\Delta_{L}(h^v_I, h^w_I)\n\\
&\equiv& \delta_{r,r'}\delta_{t^{\va v}, t^{\va w}}\delta_{s^{\va v}, s^{\va w}}
 \int_{\va SU(2)} d\gamma \prod_{I=1}^4 \delta(\gamma h^v_I (h^w_I)^{-1})\,, 
\ea
where the r.h.s. guarantees the left gauge invariance of the vertex wavefunction, namely
\begin{equation}\la{left-inv}
\hsigma_{r,t^{\va v}s^{\va v}}(h^v_I) =\hsigma_{r,t^{\va v}s^{\va v}}(\gamma h^v_I)\,,~ \forall \gamma \in \SU(2).
\end{equation} 
Moreover, the $\delta_{r,r'}$ implies that operators associated to different shells commute with each other. 
The above field operators are constructed out of the fundamental GFT field operators by convolution with the condensate wavefunction $\sigma$; they thus create/annihilate GFT quanta, all associated with such wavefunction. This association with a unique wavefunction for all the quanta forming a given shell is what we call \lq wavefunction homogeneity\rq, which puts these states in correspondence with homogeneous (continuum) spatial geometries, and also what characterizes the same states as GFT condensates. The condensate quanta are then glued to one another, for given radial parameter, to form the 3D triangulations constituting the shell, with quantum correlations encoding spatial topology \cite{Donnelly:2008vx,Chirco:2017xjb}.

In order to form a full space foliation we glue 
all the radial links belonging to the outer boundary of a given shell $r$ with the radial links belonging to the inner boundary of the shell $r+1$. For the glueing to be done consistently, the two boundaries have to have the same number and colour of radial links. Two glued shells can be graphically represented as
\be\la{Shells}
\begin{array}{c}
\includegraphics[width=3cm]{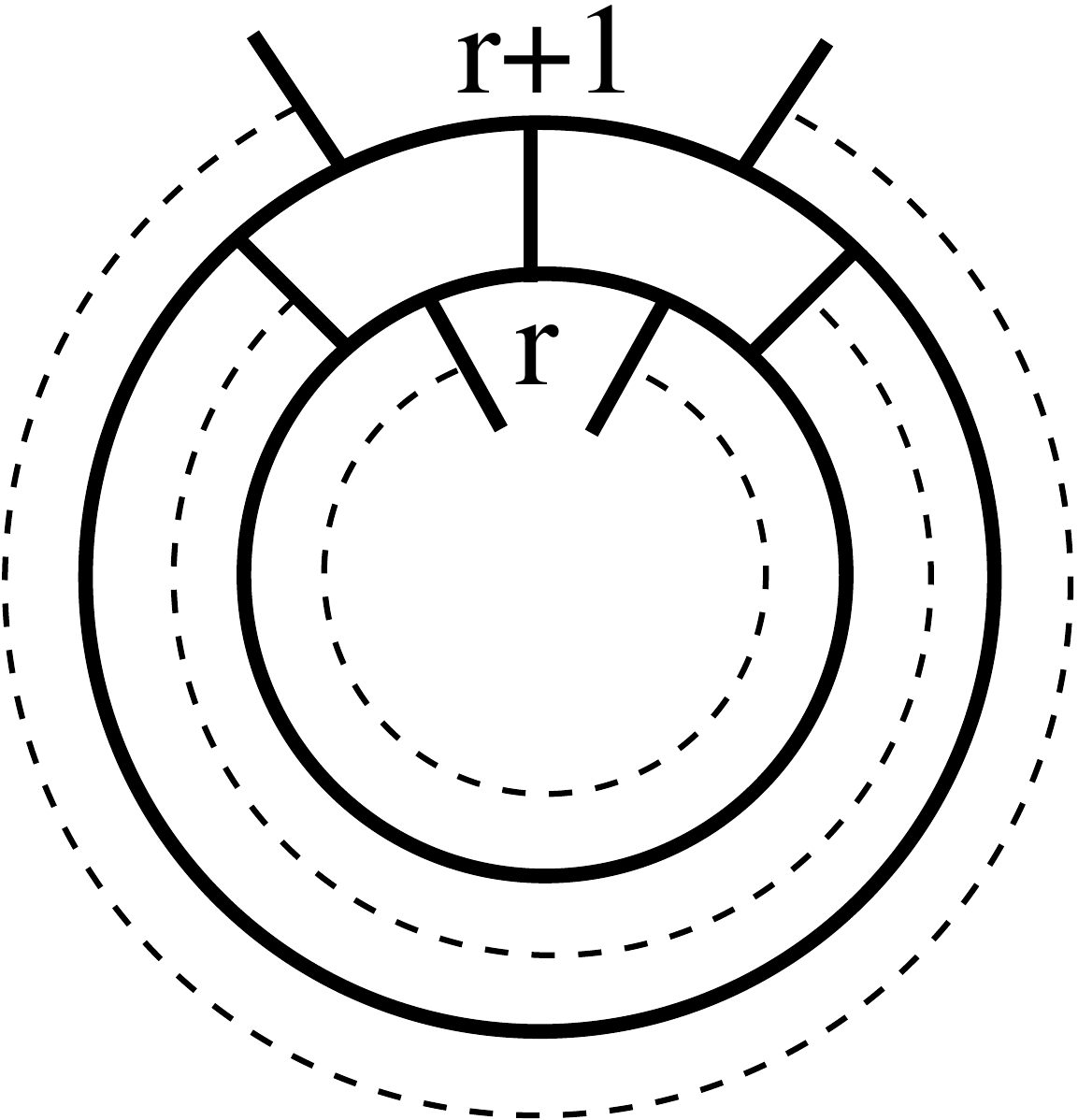}\,.
\end{array}
\ee

The use of bipartite coloured graphs in order to being able to encode the information about the spatial topology suggest that it is most convenient to adopt a construction of the seed state in terms of melonic graphs, and of the associated refinement operators in terms of dipole (or melonic) moves.
Explicitly, the seed state for a given shell $r$ is graphically represented as
\be\la{seedgraph}
\begin{array}{c}
\includegraphics[width=7cm]{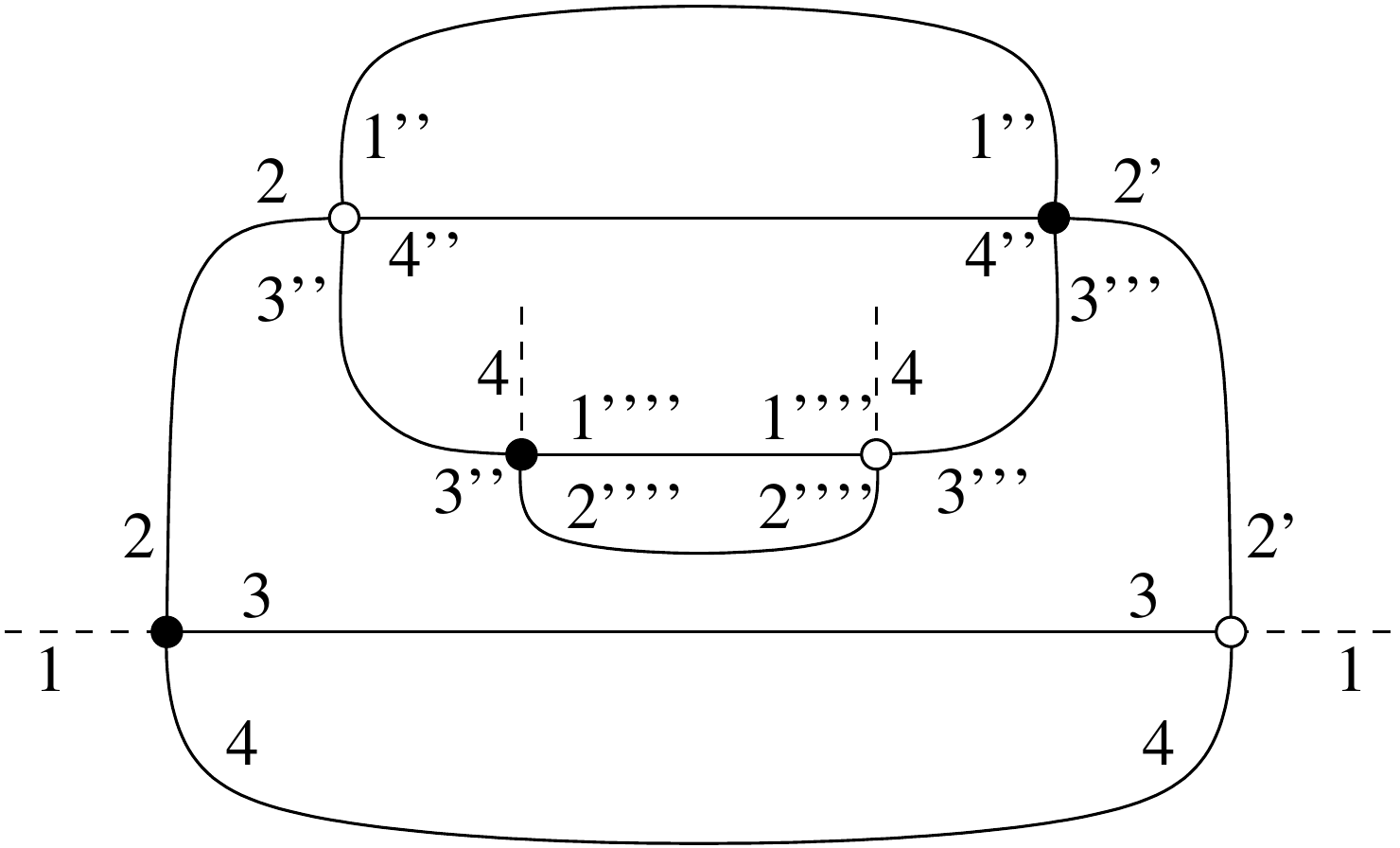}
\end{array}
\ee
and, in terms of field operators, the seed state is given by
\ba\la{tau}
\ket{\tau} &=&
\int (dg)^{10} 
\hsigmad_{\va r, B+}(e,g_{2},g_{3},g_4)
\hsigmad_{\va r,  W+}(e,g'_{2},g_{3},g_4)
\hsigmad_{\va r,  B0}(g''_{1},g'_{2},g'''_{3},g''_4)\n\\
&&~~~~~~~~~~~\hsigmad_{\va r, W0}(g''_{1},g_{2},g''_{3},g''_4)
\hsigmad_{\va r, B-}(g''''_{1},g''''_{2},g''_{3},e)
\hsigmad_{\va r, W-}(g''''_{1},g''''_{2},g'''_{3},e)
\fv \, ,
\ea
where we have arbitrarily assigned the colour 1 to the radial links of the boundary \bdyp~and the colour 4 to the boundary \bdym, and, for the moment, we have set the gluing group elements $h$'s 
associated to both sets of radial links to the identity. 

There are three refinement operators for each shell: Two refine the boundaries and one the bulk vertices. The complete set of refinement operators has been studied in \cite{Oriti:2015qva}.
Here we just concentrate on one of the two boundaries, namely the \bdyp~one. The construction for the other one follows the same logic. The action of the operator for the refinement of white vertices has a 
simple graphical representation:
\be
\move_{\va r,  W +}:~~
\begin{array}{c}
\includegraphics[width=1.8cm]{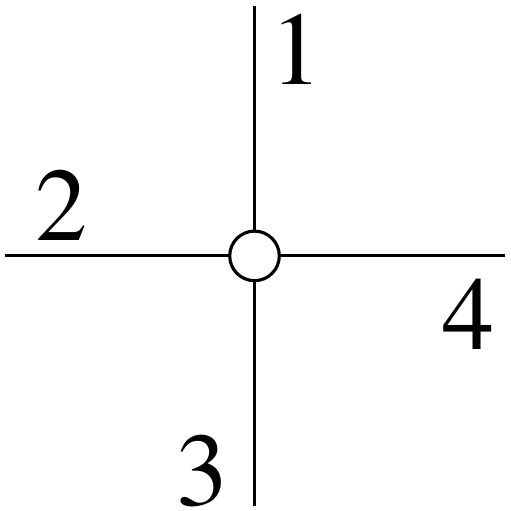}
\end{array}~~~\rightarrow~~~
\begin{array}{c} 
\includegraphics[width=5.5cm]{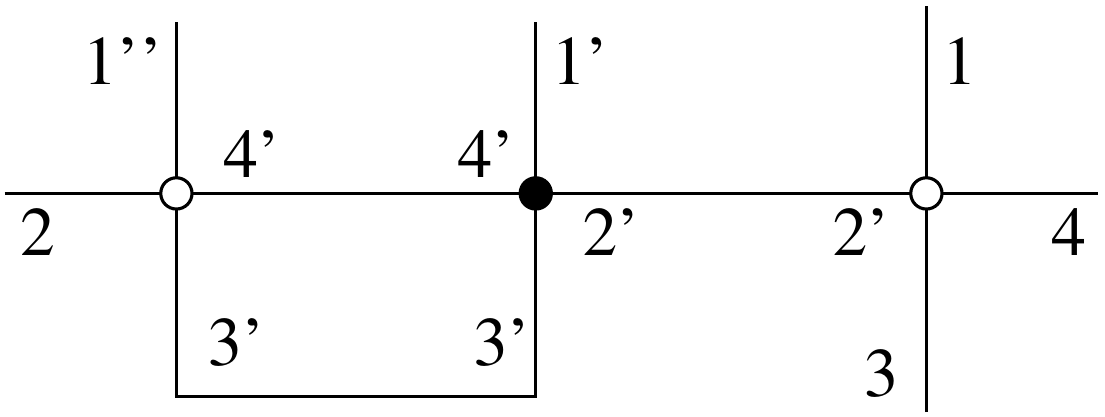}\la{refW}\,,
\end{array}
\ee
with the one for black vertices having a similar structure (notice the colours of the edge in the loop, however):
\be
\move_{\va r,  B +}:~~
\begin{array}{c}
\includegraphics[width=1.8cm]{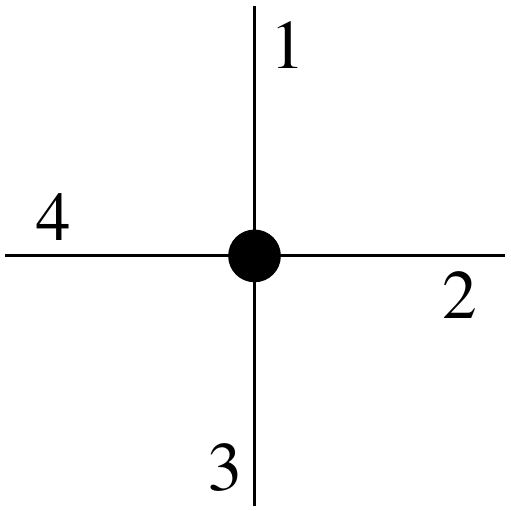}
\end{array}~~~\rightarrow~~~
\begin{array}{c}
\includegraphics[width=5.5cm]{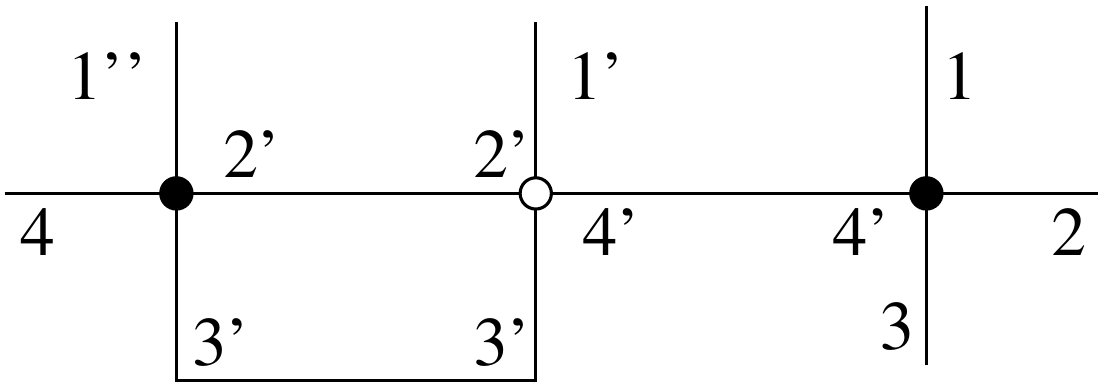}\la{refB}\,.
\end{array}
\ee
These two moves are the ones involving the minimum number of vertices, keeping fixed the topology. 
In terms of the group fields, the two move operators read
\begin{align}
\move_{\va r, W +}
\equiv \int 
&
dk_2 dk_3 dk_4 dh_{4'} dh_{2'} dh_{3'} 
\nonumber
\\ 
&
\hsigma^{\dagger}_{\va W-}(e,k_{2},h_{3'},h_{4'})
\hsigma^{\dagger}_{\va B-}(e,h_{2'},h_{3'},h_{4'})
\hsigma^{\dagger}_{\va W-}(e,h_{2'},k_{3},k_{4})
\hsigma_{\va W-}(e,k_{2},k_{3},k_{4})
\label{move2}
\end{align}
and
\begin{align}
\move_{\va r, B +}
\equiv \int 
&
 dk_2 dk_3 dk_4 dh_{4'} dh_{2'} dh_{3'} 
\nonumber
\\ 
&
\hsigma^{\dagger}_{\va r,B+}(e,h_{2'},h_{3'},k_{4})
\hsigma^{\dagger}_{\va r, W+}(e,h_{2'},h_{3'},h_{4'})
\hsigma^{\dagger}_{\va r, B+}(e,k_{2},k_{3},h_{4'})
\hsigma_{\va r,B+}(e,k_{2},k_{3},k_{4})\,.
\label{move1}
\end{align}

These are graph topology-preserving operators and their actions \eqref{refB}, \eqref{refW} can be straightforwardly verified by computing their commutators with, respectively, $\hsigmad_{\va r, B+}(e,g_{2},g_{3},g_4), \hsigmad_{\va r, W+}(e,g_{2},g_{3},g_4)$.
In a similar fashion, we can build refinement operators for vertices belonging to the other two components of the shell. 
We can thus arbitrarily refine the seed state \eqref{tau} by repeated action of the operators $\move_{\va r, t s}$ in order to implement the sum over triangulations while preserving the desired topology and the key feature of wavefunction homogeneity. In fact, the refinement moves are implemented through operators built out of field operators \eqref{c-field} dressed with the same wavefunction as the seed state; in this way, the geometric information of the finer shell states is still encoded in the same small number of parameters.
The state of a given shell $r$ can then be written as
\begin{equation}\la{shell-state}
\ket{\Psi_r} =  F_{r}(\move_{\va r,  B s},\move_{\va r, W s}) \ket{\tau}\,,
\end{equation}
where, at this stage, $F_r$ is a generic function of the refinement operators associated to the given shell $r$. 

This completes the definition of the spherically symmetric quantum states we will use in the following, for describing the microstructure of quantum black holes.

\subsection{The area operator}\la{sec:area}
The physical properties encoded in our quantum states have to be extracted by computing suitable operators. While the combinatorial aspects of the GFT formalism, shared with random tensor models, were crucial in the definition of our quantum states, alongside its 2nd quantization tools, the quantum geometric aspects, shared with loop quantum gravity (and simplicial approaches), become prominent in the physical interpretation of the same states and in the identification of interesting operators. 

An important geometric operator for our purposes is the area operator. 
Following the prescription in \cite{Oriti:2015qva}, a second quantized version of a shell boundaries area operator is defined by
\begin{equation}\la{area}
\hat{\mathbb{A}}_{Jr,s}=\sum_{t=\va {B,W}}\hat{\mathbb{A}}_{Jr,t s} \equiv \kappa \sum_{t=\va {B,W}} \int dh_I^v \hsigmad_{r,ts}(h^v_I) \sqrt{E^{i}_{J} E^{j}_{J} \delta_{ij}} \rhd \hsigma_{r,ts}(h^v_I)\,,
\end{equation}
where $\kappa=8\pi\gamma \ell_P^2$, introducing the dependence on the Barbero--Immirzi parameter $\gamma$.
In the expression above the label $s$ takes values $\{+,-\}$, accordingly to which boundary of the shell $r$ we want to compute the area of, and the index  $J$ matches the number associated to the radial links dual to the given boundary.

The action of the operator \eqref{area} is computed using the definition
\begin{equation}
E^{i}_{J} \rhd f(g_I) := \lim_{\epsilon\rightarrow 0} \iu \frac{d}{d\epsilon}
f(g_{1},\ldots, e^{-\iu \epsilon \tau^{i}}g_{J}, \ldots, g_{4})
\end{equation}
for a given function $f:SU(2)^4\rightarrow \C$. It is immediate to see that the expectation value of the area operator \eqref{area} on a shell boundary state can be written as a function
of the number of quanta and \emph{a single vertex expectation value},
\begin{equation}\la{Area-op}
\langle \hat{\mathbb{A}}_{Jr,s}  \rangle =\kappa \langle \widehat{n}_{r,s}  \rangle
\int dh^v_I dg^v_I 
\sigma_{r,s}(h^{v}_I g_I^{v})  \sqrt{E^{i}_{J} E^{j}_{J} \delta_{ij}} \rhd \overline{\sigma_{r,s}(h_I^{v} g_I^{v})}\equiv \langle \widehat{n}_{r,s}  \rangle a_{Jr,s} ,
\end{equation}
where we have defined $a_{Jr,s}$ the expectation value of the area operator on a single radial link-$J$ in the boundary $s$ of the shell $r$, and $\widehat{n}_{r,s}$ is the number operator given by
\be
\widehat{n}_{r,s}=\sum_{t=\va {B,W}}\widehat{n}_{r,ts}
=\sum_{t=\va {B,W}} \int dh^v_I\, \hsigmad_{r,ts} (h^v_I) \hsigma_{r,ts} (h^v_I)\,.
\ee
Notice that, due to the definition of the seed state and the refinement operators, after each refinement action the graphs are such that 
\be
n_{r, {\va B} s}=n_{r,{\va W} s}= \frac{n_{r,s}}{2}
\ee
always holds, where $n\equiv\langle\widehat{n}\rangle $.

An analog one-body operator can be constructed for the volume. Also in this case, the factorisation property \eqref{Area-op} holds. The structure of these expectation values should
come as natural given the structure of the wavefunction, boiling down to dimensional considerations (areas are extensive quantities) and to the fact that a single wavefunction has been used.

\section{Shell condensate state: spin representation}

Let us now introduce a `dual' spin representation for generalized GFT condensates, and a different example of condensate states constructed by the same scheme but relying on this dual representation. This will provide a useful computational toolkit and, at the same time, it allows us to circumvent the issue of non-normalisability in the kinematical Hilbert space (the Fock space) of the condensate states constructed out of the field operators \eqref{c-field}. In practice, when working in the group representation, and for the condensate states described in the previous section, a regularisation 
scheme is generally required in order to obtain finite expectation values for geometric operators like \eqref{Area-op}. On the other hand, when working in the spin representation at fixed spins, and constructing adapted condensate states using the same scheme, this  issue doesn't arise. Beside these technical advantages, we detail this dual construction for two additional reasons. First, it shows the generality of our construction and results; in fact, starting from these two basic definitions of condensate states, one can envisage new definitions combining or interpolating between the two basic ones (which play a role of possible bases for linear combinations possessing similar properties). Second, this dual spin-based construction is the one producing quantum GFT condensate states that are the closest to the quantum states customarily used in loop quantum gravity as models of quantum black holes, based on eigenstates of the area operator and thus labeled by fixed spins on the links puncturing the black hole horizon. This will facilitate comparison of our results with the LQG literature.

The spin representation follows from a straightforward Peter-Weyl decomposition of the vertex wavefunction, with the field operators \eqref{c-field} now becoming
\ba
\hat{\sigma}_{r,ts}(h_I) &=& \int dg_I \sigma_{r,s}(h_Ig_I) \hphi_ {t}(g_I) = 
\int dg_I
\sum_{\{j\}, l_{\va R},l_L}
\sigma_{r,s}^{j_1\ldots j_4 l_{\va L} l_{\va R}} \,
\iota^{j_1 j_2 j_3 j_4 l_{\va L}}_{m_1m_2m_3m_4} 
{\iota^{j_1 j_2 j_3 j_4 l_{\va R}}_{n_1n_2n_3n_4} }
\prod_{I=1}^{4}
D^{j_I}_{m_I o_I}(h_I)
D^{j_I}_{o_I n_I}(g_I)
\hphi_ {t}(g_I) 
\n\\
&=& \sum_{\{j\}, l_{\va R},l_L}\sigma^{j_1\ldots j_4 l_{\va L} l_{\va _R}} \iota^{j_1 j_2 j_3 j_4 l_{\va L}}_{m_1m_2m_3m_4} 
\hat{a}^{j_1\ldots j_4  l_{\va R}}_{(r,ts)\,o_1 \ldots o_4} 
\prod_{I=1}^{4}
D^{j_I}_{m_I o_I}(h_I)\,,
\ea
where we defined the following new field operators
\begin{equation}\la{a-field}
\hat{a}^{j_1\ldots j_4 l_{\va R}}_{(r,ts)\, o_1 \ldots o_4} 
=\sigma_{r,s}\,
 \iota^{j_1 j_2 j_3 j_4 l_{\va R}}_{n_1n_2n_3n_4} 
\int dg_I \prod_{I=1}^{4} D^{j_I}_{o_I n_I}(g_I) \hphi_ {t}(g_I) \,,
\end{equation}
and we have assumed the factorisation property 
\be
\sigma_{r,s}^{j_1\ldots j_4 l_{\va L} l_{\va R}}=\sigma_{r,s}\,\sigma^{j_1\ldots j_4 l_{\va L} l_{\va R}}\,.
\ee
One might wonder whether such a restriction is plausible or not. 
Let us notice that this particular choice means that different layers differ more because of their 
volume (controlled essentially by the number of quanta, determined by $|\sigma_{r,s}|^2$) 
rather than their intrinsic geometry (type of curvature, for instance). 
This situation is very natural in spherical symmetry, as it is its very definition in the
continuum setting: the geometry of different two-dimensional spheres singled out by the 
isometry group differs only by a rescaling determined by the radial coordinate, the rest
being constrained by the isometry group itself.

From the commutation relation  of the basic field operators
\begin{equation}\la{phi-comm}
	[\hphi_t(g_I), \hphid_{t'}(g'_I)] =\delta_{t,t'} \Delta_{R}(g, g') \equiv\delta_{t,t'} \int_{\va SU(2)} d\gamma 
	\prod_{I=1}^4 \delta(g_I \gamma (g'_I)^{-1})\,
\end{equation}
and the orthonormality
relations between the Wigner matrices
\be
\int dg D^j_{mn}(g)\overline{D^{j'}_{m'n'}(g)}=\frac{1}{d_j}\delta_{j,j'}\delta_{m,m'}\delta_{n,n'}\,,
\ee
it is immediate to see that the new field operators \eqref{a-field}
satisfy
\begin{equation}\la{a-comm}
[
\hat{a}^{j_1\ldots j_4 l_{\va R}}_{(r,ts)\,o_1 \ldots o_4} 
,
\hat{a}^{\dagger j'_1\ldots j'_4 l'_R}_{(r',t's')\,o'_1 \ldots o'_4} 
]
=|\sigma_{r,s}|^2\delta_{r,r'}\delta_{t,t'}\delta_{s,s'}
\delta_{l_R,l'_{R}} n(j_1,j_2,j_3,j_4, l_{\va R})
\prod_{I=1}^4
\frac{1}{d_{j_I}}
\delta_{o_I o'_I}
\delta_{j_I j_I'}\,.
\end{equation}
where
 \be
  \delta^{l,l'} n(j_1,j_2,j_3, j_4,l)=\sum_{\{ m \}}
\suinter{1}{2}{3}{4}{l}{m}
\suinter{1}{2}{3}{4}{l'}{m}
\ee
is the intertwiners normalization factor. With the convention 
\begin{equation}
\suinter{1}{2}{3}{4}{l}{m} = \sum_{m,m'}
C^{j_1 j_2 l}_{m_1 m_2 m}
C^{j_3 j_4 l'}_{m_3 m_4 m'}
C^{l l' 0}_{m m' 0} \,,
\end{equation}
where the $C$'s are the Clebsch-Gordan coefficients, the normalization factor $n(j_1,j_2,j_3, j_4,l)$ is equal to 1 and therefore we omit it from now on.

The wavefunction of the new field operators then reads
\be\la{a-wave}
a^{j_1\ldots j_4 l_{\va R}}_{(r,s)\, o_1 \ldots o_4}(g_I)\equiv \langle g_t|\hat{a}^{\dagger j_1\ldots j_4 l_{\va R}}_{(r,ts)\, o_1 \ldots o_4} |0\rangle
=\sigma_{r,s}\,
 \iota^{j_1 j_2 j_3 j_4 l_{\va R}}_{n_1n_2n_3n_4} 
\prod_{I=1}^{4} D^{j_I}_{o_I n_I}(g_I) \,.
\ee

\subsection{Refinement operators}

We now write the refinement operators $\move_{r, \va B s},\move_{r, \va W s}$ in terms of the spin representation field operators \eqref{a-field}. We concentrate on 
the operators for the outer boundary $s=+$. Refinement operators for the other boundaries  can be constructed straightforwardly in a similar fashion. Assuming again this boundary radial links to have label 1, we define
\ba\la{MB}
\move^{j_1j_2j_3 j_4 l_{\va R}}_{(\va r,B+)\,\,  o_1 o'_1 o''_1 o'''_1}=\frac{1}{|\sigma_{r,s}|^2}\left(\prod_{I=1}^4
{d_{j_I}}\right)\sum_{\{m\}} \hat{a}^{\dagger\,j_1j_2j_3 j_4 l_{\va R}}_{(\va r,B+)\,\,o_1 m_2 m_3 m'_4}  \hat{a}^{\dagger\,j_1j_2j_3 j_4 l_{\va R}}_{(\va r,W+)\,\,o''_1 m'_2 m'_3 -m'_4} \hat{a}^{\dagger\,j_1j_2j_3 j_4 l_{\va R}}_{(\va r,B+)\,\,o'_1 -m'_2 -m'_3 m_4}\hat{a}^{j_1j_2j_3 j_4 l_{\va R}}_{(\va r,B+)\,\,o'''_1 m_2 m_3 m_4}\,,
\ea
where notice the absence of the sign flips in the last term, due to the fact that we have $\hat a$ instead of $\hat a^\dagger$.

By means of \eqref{a-comm}, we can now verify that the refinement operators above realises the move depicted in \eqref{refB}, namely
\ba
[\move^{j_1j_2j_3 j_4 l_{\va R}}_{(\va r,B+)\,\,o_1 o'_1 o''_1 o'''_1}, \hat{a}^{\dagger\,j'_1j'_2j'_3 j'_4 l'_{\va R}}_{(\va r,B+)\,\,n_1 n_2 n_3 n_4}]=
\sum_{\{m\}} \hat{a}^{\dagger\,j_1j_2j_3 j_4 l_{\va R}}_{(\va r,B+)\,\,o_1 m'_2 n_3 n_4}  \hat{a}^{\dagger\,j_1j_2j_3 j_4 l_{\va R}}_{(\va r,W+)\,\,o''_1 -m'_2 m'_3 m'_4} \hat{a}^{\dagger\,j_1j_2j_3 j_4 l_{\va R}}_{(\va r,B+)\,\,o'_1 n_2 -m'_3 -m'_4}\delta_{o'''_1, n_1}\delta_{l_{\va R}, l'_{\va R}}\prod_I\delta_{j_I,j'_I}\,.
\ea
Similarly, we have
\ba\la{MW}
\move^{j_1j_2j_3 j_4 l_{\va R}}_{ (\va r,W+)\,\,  o_1 o'_1 o''_1 o'''_1}=\frac{1}{|\sigma_{r,s}|^2}
\left(\prod_{I=1}^4
{d_{j_I}}\right)\sum_{\{m\}}
 \hat{a}^{\dagger\,j_1j_2j_3 j_4 l_{\va R}}_{(\va r,W+)\,\,o_1 m'_2 m_3 m_4} 
 \hat{a}^{\dagger\,j_1j_2j_3 j_4 l_{\va R}}_{(\va r,B+)\,\,o''_1 -m'_2 m'_3 m'_4} 
 \hat{a}^{\dagger\,j_1j_2j_3 j_4 l_{\va R}}_{(\va r,W+)\,\,o'_1 m_2 -m'_3 -m'_4}
 \hat{a}^{j_1j_2j_3 j_4 l_{\va R}}_{(\va r,W+)\,\,o'''_1 m_2 m_3 m_4}\,,
\ea
and a direct calculation as in the black case shows that the operator above implements the action \eqref{refW}.

\subsection{Area expectation value}

Let us compute the area expectation value on a single-vertex state created by the field operator \eqref{a-field}. Consistently with our previous convention, vertex radial link dual to the boundary of interest has colour $1$. Using \eqref{Area-op}, \eqref{a-wave}, we get
 \ba
 a_{1r,s}&=&\kappa\int dg_I
  a^{j_1\ldots j_4 l_{\va R}}_{(r,s)\, o_1 \ldots o_4}(g_I)  \sqrt{E^{i}_{1} E^{j}_{1} \delta_{ij}} \rhd\overline{a^{j_1\ldots j_4 l_{\va R}}_{(r,s)\, o_1 \ldots o_4}(g_I)}\n\\
  &=&\kappa\int dg_I |\sigma_{r,s}|^2
  \iota^{j_1 j_2 j_3 j_4 l_{\va R}}_{m_1m_2m_3m_4} \iota^{j_1 j_2 j_3 j_4 l_{\va R}}_{n_1n_2n_3n_4} \sqrt{j_1(j_1+1)}
\prod_{I=1}^{4} D^{j_I}_{o_I m_I}(g_I) \overline{D^{j_I}_{o_I n_I}(g_I) }\n\\
&=&\kappa|\sigma_{r,s}|^2\sqrt{j_1(j_1+1)}\prod_{I=1}^{4}\frac{1}{d_{j_I}}\,. 
 \ea
 
As anticipated above, the only dependence on the vertex wavefunction is through the 
normalisation factor $|\sigma_{r,s}|^2$. As we have already remarked, this might seem 
suspicious but it is to be expected if we take seriously the idea that these states have leaves 
which support the same symmetry group.

\section{Shell density matrix}\la{sec:dens}

Given the single shell state \eqref{shell-state}, we can construct a pure state associated to a full 3D spatial foliation through a shell-gluing procedure.
This can be obtained through the definition of a full foliation refinement operator built out of the single-shell ones. At this kinematical stage, there is considerable freedom left
in such a definition, with the only constraint coming from the requirement of spatial topology preservation. More precisely, the refinement and gluing operations have to be carried
out in such a way that the outer boundary of a given shell $r$ always contains the same number of vertices as the inner boundary of the shell $r+1$ it is glued to, so that no open links are created. 
It should be remarked, however, that the structure of the refinement operators associated to the three different components of a given shell is such that these can act independently on one another. 
Therefore, without further inputs coming from the dynamics, the most generic complete-foliation state can be written as a product of single-shell states, namely
\begin{equation}\la{full-state}
\ket{\Psi}=\prod_r \ket{\Psi_r} \,,
\end{equation}
upon which the constraint on the `synchronisation' of nearby shell boundaries refinement is applied (thus removing the factorisation into shells and introducing correlations). 

From the pure state \eqref{full-state} we can obtain the density matrix
\be
\hat \rho=|\Psi\rangle\langle \Psi|
\ee
of the full 3D spatial foliation. 

\subsection{Reduced Density Matrix for a Single Shell}
We now want to compute the reduced density matrix associated to the outer boundary of a single shell $r$ by tracing over  the rest of the bulk state. This reduced density matrix will be the central object for our entropy calculation. 

In general, the boundary component of a given shell state \eqref{shell-state} contains a superpositions of all graphs that can be obtained by all possible combinations of strings of refinement operators associated to the boundary.  The coefficients of such superposition are determined by the specific form of the function $F_r$. We will come back to this important point in a moment. For now, in order to understand the entanglement structure between different shells,
let us start by simply considering a given graph $A$ associated to the shell $r$ outer boundary  and the graph $B$ of the inner boundary of the shell $r+1$ right outside. The result of this simple example can be easily generalised to the rest of the graph. Therefore, if we assume that both graphs are formed by $n$ vertices (in order to be properly glued they must have the same number of building blocks), where again we take the connecting radial links of colour 1, we can write the wavefunction of these two components as 
\be
\psi(g^{\va {A_1}}_I,\ldots,g^{\va {A_n}}_I,g^{\va {B_1}}_I,\ldots,g^{\va {B_n}}_I)= \prod_{i=1}^{n}
a^{j_1\ldots j_4 l_{\va R}}_{{\va A_i}\, m^i_1 \ldots m^i_4}(g^{\va A_i}_I)
a^{j_1\ldots j_4 l_{\va R}}_{{\va B_i}\, n^i_1 \ldots n^i_4}(g^{\va B_i}_I)\delta_{m^i_1,-n^{t^m_1(i)}_1}\prod_{J=2}^4\delta_{m^i_J,-m^{t^m_J(i)}_J}\delta_{n^i_J,-n^{t^n_J(i)}_J}\,,
\ee
where the $\delta$'s are used to keep track of the connectivity of the whole graph $A\cup B$, with the notation $t^m_J(i)$ $(t^n_J(i))$ indicating the target vertex  in the graph $A$ $(B)$ of the edge of colour $J$ departing from the vertex $i$ in the graph $A$ $(B)$, and similarly for $t^m_1(i)$ with the target vertex in the graph $B$ (instead of $A$), encoding the connectivity between the two boundaries through the radial links  of colour 1. We can thus write the total density matrix as
\ba
&&\rho^{(n)}(g_I^{{\va A_1}},...,g_I^{{\va A_n}},g_I^{{\va B_1}},...,g_I^{{\va B_n}};
{g'_I}^{{\va A_1}},...,{g'_I}^{{\va A_n}},{g'_I}^{{\va B_1}},...,{g'_I}^{{\va B_n}})\\
&&= C
\prod_{i=1}^{n}
a^{j_1\ldots j_4 l_{\va R}}_{{\va A_i}\, m^i_1 \ldots m^i_4}(g^{\va A_i}_I)
a^{j_1\ldots j_4 l_{\va R}}_{{\va B_i}\, n^i_1 \ldots n^i_4}(g^{\va B_i}_I)\delta_{m^i_1,-n^{t^m_1(i)}_1}\prod_{J=2}^4\delta_{m^i_J,-m^{t^m_J(i)}_J}\delta_{n^i_J,-n^{t^n_J(i)}_J}\n\\
&&
\overline{a^{j_1\ldots j_4 l_{\va R}}_{{\va A_i}\, {m'}^i_1 \ldots {m'}^i_4}({g'}^{\va A_i}_I)
a^{j_1\ldots j_4 l_{\va R}}_{{\va B_i}\, {n'}^i_1 \ldots {n'}^i_4}({g'}^{\va B_i}_I)}\delta_{{m'}^i_1,-{n'}^{t^{m'}_1(i)}_1}\prod_{J=2}^4\delta_{{m'}^i_J,-{m'}^{t^{m'}_J(i)}_J}\delta_{{n'}^i_J,-{n'}^{t^{n'}_J(i)}_J}\,,
\ea
where $C$ is a normalisation factor. 

If we now use the relation
\begin{equation}\la{delta-spin}
\int dg_I \,a^{j_1\ldots j_4 l_{\va R}}_{(r,s)\, m_1 \ldots m_4}(g_I)\overline{a^{j_1\ldots j_4 l_{\va R}}_{(r,s)\, n_1 \ldots n_4}(g_I)}= 
 |\sigma_{r,s}|^2
\prod_{I=1}^4
\frac{1}{d_{j_I}}
\delta_{m_I, n_I}\,,
\ee
following from the commutation relation \eqref{a-comm}, to integrate away the $B$ part, we get
\ba\la{red-spin}
\rho^{(n)}_{A}(g_I^{1},...,g_I^{n};{g'_I}^{1},...,{g'_I}^{n})&=&
\left(  \frac{\prod_{I=1}^4 d_{j_I}}{|\sigma_{r,s}|^2}\right)^n
\prod_{i=1}^{n}
a^{j_1\ldots j_4 l_{\va R}}_{{\va A_i}\, m^i_1 \ldots m^i_4}(g^i_I)
\overline{a^{j_1\ldots j_4 l_{\va R}}_{{\va A_i}\, {m'}^i_1 \ldots {m'}^i_4}({g'_I}^i)}\n\\
&\times&\delta_{m^i_1,{m'}^i_1}
\prod_{J=2}^4\delta_{m^i_J,-m^{t^m_J(i)}_J} \,
\delta_{{m'}^i_J,-{m'}^{t^{m'}_J(i)}_J}\,,
\ea
where we have set the normalisation factor $C=\left( \prod_{I=1}^4 d_{j_I}/|\sigma_{r,s}|^2\right)^{2n}$.
We see that the normalised reduced density matrix we obtain is mixed, as  a consequence of the relation $m^i_1={m'}^i_1$ imposed by the first set of $\delta$'s and following from the property \eqref{delta-spin}. 

Remarkably, as it emerges from this simple example, any property about the rest of the graph traced away disappears: different completions of the same visible portions of the state will lead to the same reduced density matrix. 
This is a direct consequence of the commutation relation \eqref{a-comm} (which implies \eqref{delta-spin}). This holographic property of our states will remain valid also when tracing away a bigger graph in the bulk, exactly for the same 
mechanism. This means that, given a graph for the whole 3D space foliation, if we choose the boundary of an arbitrary shell $r$ and trace away all the rest of the graph, we end up with a mixed 
reduced density matrix which  contains {\it no information} about the bulk degrees of freedom. The only entanglement that remains in the reduced density matrix is the one induced by the radial links of 
the closest shell. 
Let us reiterate the message: for these particular states, all the rest of the information about the remaining graph disappears from the reduced density matrix, regardless of how big or intricate the rest of the graph is. This means that the 
entanglement entropy contribution comes uniquely from the entanglement with the closest shell, as expected by standard calculation in QFT on a sphere \cite{Solodukhin:2011gn}.

\subsection{Entanglement Entropy}\la{sec:eentropy}
To compute the entanglement entropy we now need to diagonalise the reduced density matrix, that is, we need to compute the eigenvectors of \eqref{red-spin}. Let us consider the state
\be\la{eigenstates-spin}
\Psi^{(n)}_A(n_1, {g})
=\left( \frac{\prod_{I=1}^4d_{j_I}}{|\sigma_{r,s}|^2}\right)^{n/2}
\prod_{i=1}^{n}
\overline{a^{j_1\ldots j_4 l_{\va R}}_{{\va A_i}\, n^i_1 \ldots n^i_4}(g_I^i)}
\prod_{J=2}^4\delta_{n^i_J,-n^{t^n_J(i)}_J}\,,
\ee
which satisfies
\be
\langle \Psi^{(n)}_A(n_1, {g})| \Psi^{(n)}_A(n'_1, {g})\rangle= \prod_{i=1}^n \delta_{n^i_1, n'^i_1}\,,
\ee
and compute
\ba
&& \int \prod_{i=1}^{n} dg_I^{i}\rho^{(n)}_{red}(g_I^{{1}},...,g_I^{{\n}};{g'_I}^{{1}},...,{g'_I}^{{n}}) \Psi^{(n)}_A(n_1, g)\n\\
&& =\left(  \frac{\prod_{I=1}^4 d_{j_I}}{|\sigma_{r,s}|^2}\right)^{\frac{3}{2}n}\int \prod_{i=1}^{n} dg_I^{i}
a^{j_1\ldots j_4 l_{\va R}}_{{\va A_i}\, m^i_1 \ldots m^i_4}(g^{i}_I)
\overline{a^{j_1\ldots j_4 l_{\va R}}_{{\va A_i}\, {m'}^i_1 \ldots {m'}^i_4}({g'_I}^{i})}
\delta_{m^i_1,{m'}^i_1}
\prod_{J=2}^4\delta_{m^i_J,-m^{t^m_J(i)}_J} \,
\delta_{{m'}^i_J,-{m'}^{t^{m'}_J(i)}_J}\n\\
&&\times\, \overline{a^{j_1\ldots j_4 l_{\va R}}_{{\va A_i}\, n^i_1 \ldots n^i_4}(g^{i}_I)} 
\prod_{J=2}^4\delta_{n^i_J,-n^{t^n_J(i)}_J}\n\\
&&=\left(  \frac{\prod_{I=1}^4 d_{j_I}}{|\sigma_{r,s}|^2}\right)^{\frac{n}{2}}
\overline{a^{j_1\ldots j_4 l_{\va R}}_{{\va A_i}\,m^i_1 \ldots {m}^i_4}({g'_I}^{i})}
\prod_{J=2}^4
\delta_{{m}^i_J,-{m}^{t^{m}_J(i)}_J}\n\\
&&= \Psi^{(n)}_A(m_1, {g'})\,,
\ea
where we have used the relation  \eqref{delta-spin} again.
From the calculation above, we see that the states \eqref{eigenstates-spin}
are eigenstates of the shell reduced density matrix \eqref{red-spin} with eigenvalue $1$. 

We can thus label all the eigenstates of the shell reduced density matrix using a graph basis and  the notation $\Psi_{r,s}^{(n)}(\Gamma_\alpha)$ to denote the states \eqref{eigenstates-spin}, 
where the structure of the given graph $\Gamma_\alpha$ is encoded in the product of deltas $\prod_{J=2}^4\delta_{n^i_J,-n^{{}^\alpha t^n_J(i)}_J}$.

Given that this does not hold in general \cite{Oriti:2014uga}, it is instructive to show explicitly the orthogonality of the states $\Psi_{r,s}^{(n)}(\Gamma_\alpha)$ for different graphs $\Gamma_\alpha, \Gamma_{\alpha'}$. We want to prove that  
\be\la{ortho}
\langle \Psi_{r,s}^{(n)}(\Gamma_\alpha) |\Psi_{r,s}^{(n)}(\Gamma_{\alpha'}) \rangle=\delta_{\alpha, \alpha'}\prod_{i=1}^n \delta_{n^i_1, n'^i_1}\,.
\ee
To do so, for all $\alpha\neq \alpha'$, it is enough to consider one simple case, namely graphs with $n_{\va B}=n_{\va W}=n/2=2$ (recall that all the graphs have always the same number of black and white vertices). These states are created by acting once with the refinement operators \eqref{refW}, \eqref{refB} on the seed state for the shell boundary
\be\label{seed}
\includegraphics[width=6cm]{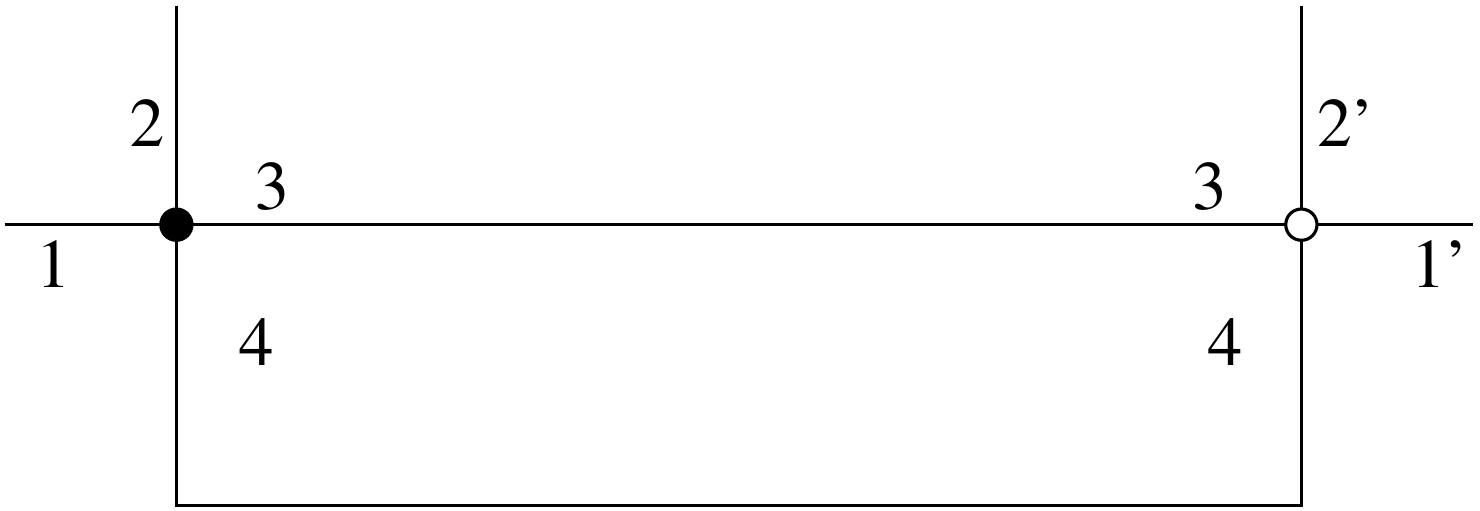}\,,
\ee
where links-1 are the radial ones and links-2 are connected to the bulk of the shell. 
The two possible states with 4 vertices, when acting with the black \eqref{refB} and white \eqref{refW} refinement operators respectively,  thus are
\ba\la{Psi1}
&&|\Psi^{(4)}_{r,1}\rangle:~~
\begin{array}{c}
\includegraphics[width=7.5cm]{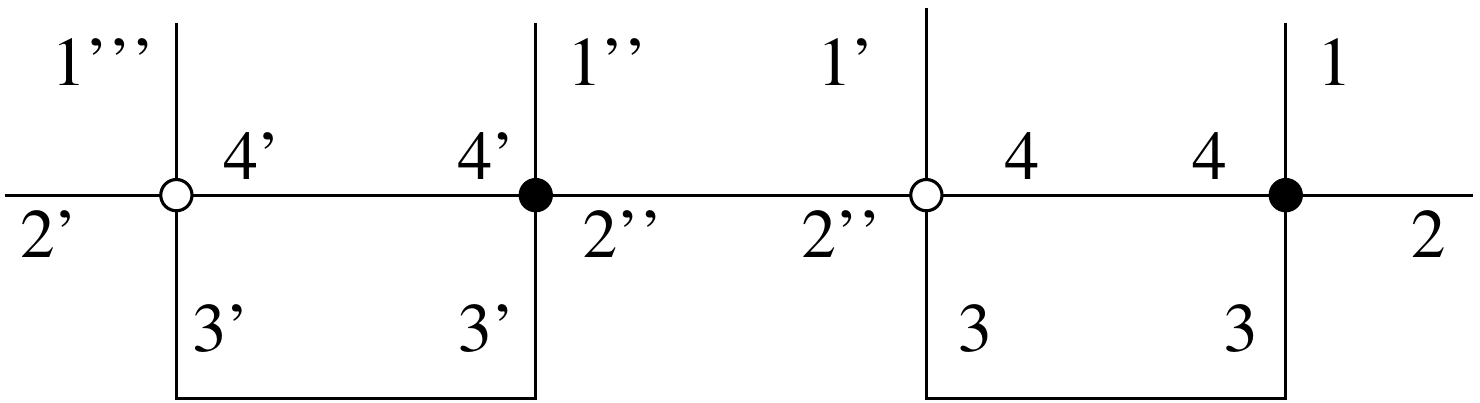}
\end{array}\\
&& |\Psi^{(4)}_{r,2}\rangle:~~
\begin{array}{c}
\includegraphics[width=5.7cm]{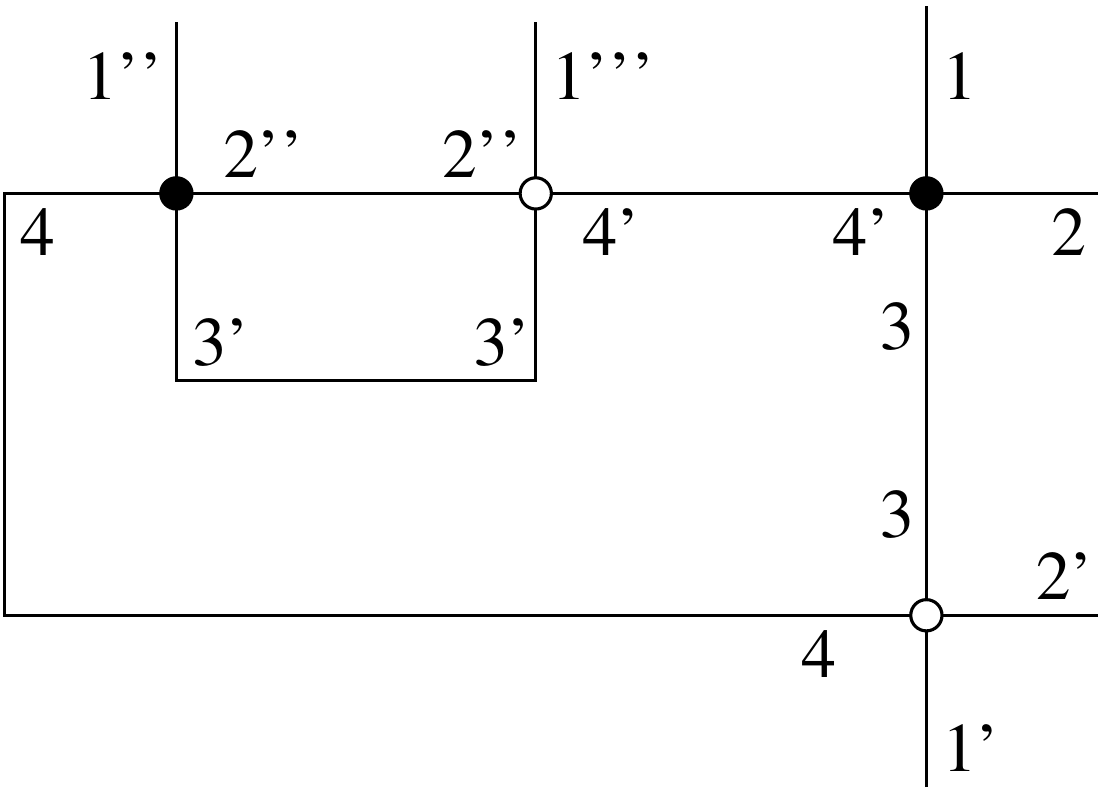}
\end{array}\la{Psi2}\,.
\ea
The scalar product between these two states gives:
\baa
\langle \Psi^{(4)}_{r,2}|\Psi^{(4)}_{r,1}\rangle&=&
\int dg^1_Idg^2_Idg^3_Idg^4_Idg'^1_Idg'^2_Idg'^3_Idg'^4_I\\
&&\overline{a^{j_1\ldots j_4 l_{\va R}}_{ m^1_1 m^1_2 m^1_3  m^1_4}(g_I^1)}
\overline{a^{j_1\ldots j_4 l_{\va R}}_{ m^2_1 m^2_2 m^2_3  -m^1_4}(g_I^2)}
\overline{a^{j_1\ldots j_4 l_{\va R}}_{ m^3_1 -m^2_2 -m^2_3  m^3_4}(g_I^3)}
\overline{a^{j_1\ldots j_4 l_{\va R}}_{ m^4_1 m^4_2 -m^1_3  -m^3_4}(g_I^4)}
\\
&& {a^{j_1\ldots j_4 l_{\va R}}_{ n^1_1 n^1_2 n^1_3  n^1_4}({g'_I}^1)}
{a^{j_1\ldots j_4 l_{\va R}}_{ n^2_1 n^2_2 -n^1_3  -n^1_4}({g'_I}^2)}
{a^{j_1\ldots j_4 l_{\va R}}_{ n^3_1 -n^2_2 n^3_3  n^3_4}({g'_I}^3)}
{a^{j_1\ldots j_4 l_{\va R}}_{ n^4_1 n^4_2 -n^3_3  -n^3_4}({g'_I}^4)}\\
&&\langle0|\hphi_ {\va W}({g'_I}^1) \hphi_ {\va B}({g'_I}^2) \hphi_ {\va W}({g'_I}^3) \hphi_ {\va B}({g'_I}^4)
 \hphid_ {\va B}(g^1_I) \hphid_ {\va W}(g^2_I) \hphid_ {\va B}(g^3_I) \hphid_ {\va W}(g^4_I)|0\rangle\\
&=& \left(\prod_{I=1}^4\frac{1}{d_{j_I}}\right)^{4} 
\Big(\delta_{m^1_1,n^1_1}\delta_{m^1_2,n^1_2}\delta_{m^1_3,n^1_3}\delta_{m^1_4,n^1_4}
\delta_{m^3_1,n^3_1}\delta_{m^2_2,n^2_2}\delta_{-m^2_3,n^3_3}\delta_{m^3_4,n^3_4}\\
&&\phantom{\prod_{I=1}^4\frac{1}{d_{j_I}}}+
\delta_{m^1_1,n^3_1}\delta_{m^1_2,-n^2_2}\delta_{m^1_3,n^3_3}\delta_{m^1_4,n^3_4}
\delta_{m^3_1,n^1_1}\delta_{-m^2_2,n^1_2}\delta_{-m^2_3,n^1_3}\delta_{m^3_4,n^1_4}\Big)\times\\
&&\phantom{\prod_{I=1}^4\frac{1}{d_{j_I}}}\Big(\delta_{m^2_1,n^2_1}\delta_{m^2_2,n^2_2}\delta_{m^2_3,-n^1_3}\delta_{m^1_4,n^1_4}
\delta_{m^4_1,n^4_1}\delta_{m^4_2,n^4_2}\delta_{m^1_3,n^3_3}\delta_{m^3_4,n^3_4}\\
&&\phantom{\prod_{I=1}^4\frac{1}{d_{j_I}}}+\delta_{m^2_1,n^4_1}\delta_{m^2_2,n^4_2}\delta_{m^2_3,-n^3_3}\delta_{m^1_4,n^3_4}
\delta_{m^4_1,n^2_1}\delta_{m^4_2,n^2_2}\delta_{m^1_3,n^1_3}\delta_{m^3_4,n^1_4}
\Big)\,,
\eaa
where we have performed the Wick contractions and used \eqref{phi-comm}. If we now expand
 this last expression, we end up with four products of $\delta$'s which all give rise to either one 
 of the following relations
\be
\delta_{m^1_3,-m^2_3}\,,~~~
\delta_{m^1_4, m^3_4}\,.
\ee
None of these two relations can be satisfied, though, due to the structure of the graph of the state \eqref{Psi2}. In fact, they would both change the shell topology by creating two disconnected regions in the boundary graph. Therefore, the scalar product vanish and the states \eqref{Psi1}, \eqref{Psi2} are  orthogonal. 
Due to the local action of the refinement operators, similar relations would follow when computing the scalar product between any two eigenstates at given $n$ corresponding to the action of a sequence of refinement operators generating two different graphs. 
This implies that all the eigenvectors $\Psi_{r}^{(n)}(\Gamma_\alpha)$ of the total reduced density matrix \eqref{red-tot-spin} are orthogonal for $\alpha\neq \alpha'$. A similar calculation shows that when $\alpha=\alpha'$ \eqref{ortho} is again satisfied. 

This seemingly unremarkable results has a very important consequence, namely
\be\la{eigenvalues-spin}
\rho^{(n)}_{r,s}(\Gamma_\alpha)\Psi_{r,s}^{(n)}(\Gamma_{\alpha'}) = \begin{cases}
   \Psi_{r,s}^{(n)}(\Gamma_{\alpha'})\quad\text{if}\;\alpha=\alpha'\\
   \\
    0\,\,\,\quad\text{if}\;\alpha\neq \alpha'\,.\end{cases}  
\ee
which implies that we can diagonalise the reduced density matrix, since we have discovered its diagonal form in terms of graphs, even in the case in which the
full sum over triangulations is kept. This result also shows that the computation of the entanglement entropy, \emph{at least in the particular corner of the Fock space
that we are exploring}, becomes a classic `counting of graphs' problem.

\section{Semi-classicality conditions}\la{sec:class}

Our construction so far is general, in the sense that it can apply to any of the shells in the space-like hypersurface of our foliation, and thus to generic spherically symmetric geometries. We now want to restrict our attention to an horizon 2-sphere cross section, \ie we want to be able to interpret one of our shells as defining a spherical {\it horizon}, and for this we need to specify horizon boundary conditions.
As we recalled in the Introduction, in the standard LQG  black hole entropy calculation these boundary conditions are specified by the notion of Isolated Horizon and they are 
implemented in the quantum theory through the imposition of the boundary conditions \eqref{IH-cond} (though most often this is already implemented at a classical level).

In the context of the GFT condensate for a spherically symmetric shell, imposition of the boundary condition \eqref{IH-cond} amounts to a relation between the flux variable $E_1$ associated to the $r$
radial link  (in the convention adopted so far for the outer boundary of a shell) of a given vertex of the shell and the holonomy around it, constructed out of the group elements associated to the 
orthogonal remaining links. The latter correspond to the group elements $g_2, g_3, g_4$ of the fundamental field operator $\hphi(g_I) $. In fact, these are the group elements which have the 
geometrical interpretation of parallel transport from the centre of a given tetrahedron to  its faces lying on the horizon. The identification of the horizon shell could then be implemented through the 
construction of a second quantised curvature operator around a given  puncture to be interpreted as the GFT holonomy operator around the radial link-1 of the corresponding vertex. However, given 
the fundamentally discrete nature of the GFT formalism and the lack of continuum manifold structures to be used as auxiliary tools to define quantum operators related to curvature, which generically involve correlation across several  graph vertices and which correspond to intensive quantities from the GFT many-body perspective, their precise definition for GFT condensates remains one of the main open technical challenges of this second quantisation formalism.

Given this obstruction, we are not going to explicitly impose IH boundary conditions in operatorial terms. 
Instead, we are going to rely on a maximum entropy argument in order to characterise the \emph{most generic} 
horizon shell geometry, as well as to capture some aspects of the semi-classical dynamics in
our so far purely kinematical construction. Such treatment of the horizon semi-classical regime seems 
natural in light of the laws of black hole thermodynamics  \cite{Bardeen:1973gs} and arguments for 
entropy bounds  \cite{Bekenstein:1980jp, Bousso:1999xy}.

The isolated horizon formalism will still play a role in our entropy calculation. In fact, we will check the compatibility 
of the single vertex Hilbert space degeneracy obtained through extremisation of the global horizon entropy with the 
constraint \eqref{IH-cond}, in the semi-classical limit. Moreover, we will also require consistency with the 
thermodynamical properties of isolated horizons. 

In addition to the notion of typicality, encoded in the maximum entropy argument,
there are other geometrical requirements to be demanded in order to guarantee
the validity of a semi-classical regime, namely: 
\begin{enumerate}
\item large horizon shell boundary area;
\item small horizon shell bulk volume;
\item small fluctuations of the horizon shell geometrical operators.
\end{enumerate}
This last requirement is guaranteed by the factorisation property of the one-body geometrical operators
(like area and volume, see Section \ref{sec:area}) in the large $n_{r,s}$ limit; from here on we thus take
$n_{r_0,s}\gg 1$, with $r_0$ denoting the horizon shell. It also follows rather generically from the condensate nature of our quantum states.

\subsection*{Holographic aspects of GFT black hole condensates}

Before proceeding with the black hole entropy calculation we need to identify the relevant degrees of freedom which contribute to it. 
In order to understand better the features of the quantum states we have constructed and of their reduced density matrix, we recall here some basic aspects of the holographic principle which is implemented by it in a specific manner.
 
We can identify two notions of holography: strong and weak \cite{Smolin:2000ag}.
In the `strong' version, the holographic principle becomes a fundamental feature of quantum gravity, 
asserting that all bulk degrees of freedom can be encoded on a boundary screen.
This is the point of view adopted  for instance in the AdS/CFT correspondence, when interpreted as an exact duality. 

There are various arguments, however, to believe that in
a background independent quantum gravity approach only a weaker notion of holography can survive \cite{Smolin:2000ag}. 
In particular, in a thermodynamical context where interactions between subsystems play a fundamental role,
the causal structure of a black hole diagram strongly suggests that the only relevant degrees of freedom to account
for the atomistic description beyond macroscopic variables are those living in the proximity of the horizon; in fact these
are the only ones in causal contact with the exterior subsystem and relevant for the analysis of an external stationary observer.
For these reasons, we are going to invoke a weak holographic principle in order to select the relevant reduced density matrix
for the black hole entropy calculation. 

Let us recall that in Section \ref{sec:dens} we modelled, before dynamical considerations are eventually
implemented, the 3D foliation of a spherically symmetric black hole geometry as a GFT pure condensate state factorised
as a product over components of different shells. In particular, this means that refinement operators could act independently
on different shells, suggesting that indeed strong holography is not realised at the microscopic level, at
least in the kinematical setting. Selection of the microscopic degrees of freedom at the origin of the statistical mechanical
nature of the Bekenstein--Hawking area entropy law thus requires us to trace out all bulk shells, both in the exterior
and the interior of the horizon shell $r_0$. 

Entanglement entropy calculations in QFT on a fixed classical background might lead to the
expectation that it should be enough to trace out only part, but not all, of the bulk Hilbert space in order to obtain
a result that scales with the area of the entangling surface \cite{Solodukhin:2011gn}. However, this expectation is borne out of examples
in which gravity is treated classically and entropy is associated to matter degrees of freedom, not gravitational ones.
As we have argued above, in a general full quantum gravitational regime holography has probably a weaker nature.

At the same time, one could consider a subclass of the GFT condensate states we constructed in which nonlocal
correlations between different shells are introduced by the imposition of some effective dynamics, which would
allow an entropy counting compatible with a stronger form of holography. The extreme case of such a scenario
would be physical states for which the refinement operators on a given shell are synchronized with those acting
on all the other shells at the same time. In this case the number of boundary graphs would exhaust all the possible
configurations of bulk graphs as well: Each boundary graph is in correspondence with just one
graph in all the rest of the bulk and there is no extra degeneracy. Such states are in fact rather straightforward to construct.

Before proceeding, a few remarks are in order.
At this stage, there is no indication that such specific sector is the physically relevant one as this would be ultimately
dictated by the quantum equations of motion, even considered in some approximate form.
The counting discussed in what follows, obtained by tracing over all the bulk vertices outside of the $r_0$ shell, 
applies straightforwardly to this particular case of strong holography.

Let us also point out that implementation of the weak holographic principle does not
automatically yield an area law for the entropy. In fact, \emph{a priori} the horizon reduced density
matrix could retain information about the rest of the bulk. It is therefore a non-trivial property
of our construction the fact that, as seen above, by tracing out bulk degrees of freedom,
the resulting reduced density matrix of the horizon is mixed but loses all the information
about the bulk beyond its existence.

In addition, as we will see in a moment, there is also a further holographic
feature concerning the degeneracy associated to the space of wavefunctions
for the single vertex Hilbert space.
This characteristic of our states is what we would expect from a causal barrier, suggesting 
that the foliation of space provided by our shell graphs could be naturally used to
mimic a foliation by null surfaces. This does not obviously represent the most general 
case of foliation, but it suits very well to the description of a black hole horizon. 
Moreover, if we were in a more general case where information about the bulk had not been washed away by the tracing operation, the implementation of weak holography for 
the horizon density matrix would have required some further restriction on the bulk 
wavefunctions, for instance coming from the dynamics. 
Due to the holographic nature of our particular states, such restriction 
is automatically included. What remains to be 
checked is the compatibility of these holographic states and their associated 
geometrical data with the imposition of the dynamics. This is far from trivial, as we have repeatedly stressed, and it is
left for future work.

Finally, it is also remarkable that we could find an orthogonal set of eigenvectors for the 
horizon density matrix \eqref{red-tot-spin}. This allows us to diagonalize it and then, due to 
the property \eqref{eigenvalues-spin}, the calculation of the horizon von Neumann entropy 
can be performed precisely, without neglecting any entanglement contribution, and this corresponds exactly to a statistical counting \emph{\'a la} Boltzmann. We have thus proven that {\it the 
horizon entanglement entropy is the same as the horizon Boltzmann entropy}. Let us 
now perform the counting.

\section{Combinatorial Contribution: Graph Counting}

Our states are defined via iterated action of refinement moves on 
the seed of each shell. The generic parametrization \eqref{shell-state}  makes use
of a generic function of two sets $(B, W)$ of refinement operators.
Written more explicitly:
\begin{equation}
\left(
\sum_{n=0}^{\infty} \prod_{s}\prod_{m=1}^n
\left( 
a^{n,m}_{\va r_0,  B s}\, \move_{\va r_0,  B s}
+
a^{n,m}_{\va r_0, W s} \,\move_{\va r_0, W s}
\right) 
\right)
\ket{\tau}
\,
.
\end{equation}
The refinement move operators, together with the seed state, specify the general 
class of possible microscopic combinatorial structures, \ie the
specific portion of the Fock space, that we are exploring. 
The parameters $a$ control instead the relative weights of the different possible
strings of refinement move operators that can be applied to the seed state.
As we have said, we will determine them following some global considerations,
making use of a maximum entropy principle, appealing to typicality of the state given
some mild general constraints.

Let us clarify this point further.
From the perspective of macroscopic, large scale dynamics, making reference to the detailed
combinatorial structure of the state does not seem a plausible approach.
Instead, the state has to be determined as the most general state compatible
with global constraints on topology, symmetry and semi-classicality.
A statistical approach based on typicality and maximum ignorance seems to be more suitable,
exploring as much as possible the space of states that we have described so far, and
limiting the truncations to specific sectors of the GFT kinematical space to the ones that respect mild constraints associated to the physical regime we are interested in.

In section \ref{sec:eentropy} we have studied the properties of the reduced density matrix, showing
a basis of states diagonalising it. We showed how the problem of computing the entropy boils down
to a (weighted!) count of graphs, but we did not proceed further in the counting.
We will now resume the computation of the entropy by first evaluating the contribution
arising from the graph proliferation.

The action of the refinement moves, via the Wick's theorem, generate a linear superposition
of states possessing different combinatorial structure, according to each particular
sequence of Wick contractions. Let $G_n$ be the set of graphs
that can be obtained via any sequence of $n$ refinement moves per shell component starting from the given
seed. The particular function in \eqref{shell-state}  will
then specify a set of weights $w_n(\Gamma)$, the coefficients entering the
diagonal of the reduced density matrix. The detailed
dependence of these weights on the coefficients $a$'s is not necessary, but
could be computed, in principle. 
Let us stress that the assumption that at each step a given horizon component
gets refined with the same number of vertices as the other two is here made
mainly to simplify the notation and it could be relaxed without affecting the final
result of the entropy counting.

We have already shown that, for a given boundary graph, the reduced density matrix 
of the horizon takes the form \eqref{red-spin}, where $A=r_0$ corresponds to the horizon shell, 
including all its three components. As argued, this form remains unchanged even when we
trace away the whole bulk degrees of freedom (both interior and exterior ones). 
Therefore, we can write the total normalized reduced density matrix of the shell for a given 
number $n$ of boundary vertices as
\be\la{red-tot-spin}
\rho_{tot, r_0}^{(n)}= \sum_{\alpha=1}^{\mathcal N} w_n(\Gamma_\alpha)  \rho^{(n)}_{r_0}(\Gamma_\alpha)\,,
\ee
where ${\mathcal N}=\#G_n$ is the total number of shell graphs for given number of vertices 
(which are $2n$, given the presence of two colours for the vertices),
obtained through all the possible actions of the refinement operators. 
Here, $\rho^{(n)}_{r_0}(\Gamma_\alpha)$ is the reduced density matrix in \eqref{red-spin}.

The presence of several graphs (at fixed topology) 
results into a combinatorial contribution to the entropy:
\begin{equation}
S_{comb} = -  \sum_{\alpha=1}^{\mathcal N} 
w_n(\Gamma_\alpha) 
\log\left( 
w_n(\Gamma_\alpha) 
\right)\,.
\end{equation}
As explained in Section \ref{sec:class}, we determine the weights $w_n$ (and then, implicitly, the function $F_r$ in \eqref{shell-state})
by maximising this entropy. It is immediate to do so: 
the most disordered configuration is the one in which the 
weights are all equal:
\begin{equation}
w^{\max}_n(\Gamma) = 
\frac{1}{{\mathcal N}}
,\qquad
S_{comb}^{\max} = \log\left( 
{\mathcal N}
\right)
\end{equation}
Therefore, the only thing left to determine is the size of the set of graphs that
can be obtained from the given set of refinement moves.

The counting of the number of graphs generated by our refinement moves can
be easily performed using familiar techniques. 
Let us focus on a given layer, for instance the outer boundary of the horizon shell. Following 
the convention adopted so far, let us suppress for a moment the edges of colour 1, whose only 
role is to connect the horizon shell with the next one.
They do not play a role in the counting since the state is completely determined by the 
specification of the combinatorial pattern of the gluings of edges of colours $2, 3, 4$. This
effectively reduces the counting problem to a counting of graphs obtained connecting 
3-valent vertices (\ie a lower dimensional problem, as it should be expected).

At a first inspection, following the reasoning in \cite{Bonzom:2011zz}, one would expect
the counting of the boundary states to give us the Catalan numbers (for D = 2, in
fact), as what we are looking at is the refinement of a melonic graph with the insertion of 
melons.
However, some care is due in the case of the shell boundaries. In fact, we have only two 
moves at our disposal, inserting loops with colours (2,3) and (3,4) only (see \eqref{refW}, 
\eqref{refB}). The insertion of loops with colours (2,4) would cause a crossing with an
edge of colour $1$. Therefore the strategy of \cite{Bonzom:2011zz} has to be slightly adapted.

Notice that the refinement moves can be seen as the insertion of melons on links of 
colour 2 and 4 only (this corresponds to the addition of tetrahedra incident on the same dual 
edge, \ie the increase of the curvature around that edge).
Therefore, our problem can be seen as the calculation of the number of ways in which
we can insert, into a line of colour 2 (for instance), a string of pairs of nodes connected
according to our rules. 

This is easily done using generating functions. 
Let $u$ be a real (or complex) parameter such that the Taylor coefficients of $G(u)$ in zero, 
which we wish to determine, are the number of graphs obtained acting with our
moves, \ie the $n$-th coefficient is the number of graphs with $n$ pairs of nodes.
The basic building blocks will be 1PI graphs, which are then combined to get all
the graphs.

The equation for the 1PI part can be obtained by inspection of the diagrammatics.
We can obtain a recursion relation observing that, due to the peculiar connectivity, the only 1PI 
diagrams are the ones in which the first black vertex and the last white vertex are connected by 
a line of colour 3. If this were not the case, one could cut the graph into two disconnected 
components removing the link between the white vertex connected with the first black vertex 
and the next black vertex.
Let us call $\Sigma(u)$ the function that generates the 1PI graphs.
We have
\begin{equation}
G(u) = \frac{1}{1+\Sigma(u)}\, .
\end{equation}
However, it is easy to realize that $\Sigma(u) = uG(u)$, as the graphs that we would be
counting are determined by all the possible ways of inserting $n-1$ pairs of nodes
in the link that would connect the first and last node of the 1PI graph. The $u$ prefactor
keeps track of the initial pair to be inserted.
Therefore:
\begin{equation}
G(u) = \frac{1}{1+u G(u)}\,.
\end{equation}
The solution, differentiable in 0, is
\begin{equation}
G(u) = \frac{-1 + \sqrt{1+4u}}{2u}\,,
\end{equation}
which, as indeed expected, is the generating function for the Catalan numbers, namely
\begin{equation}
C_{k} = \frac{(2k)!}{(k+1)! k!}\,.
\end{equation}
Therefore, the number of distinct graphs with obtained with $n$ refinement moves
acting on $l$ layers of the shell initial seed state is:
\begin{equation}\la{G}
{\mathcal N} = \left( \frac{(2n-2)!}{n! (n-1)!} \right)^l\,.
\end{equation}
As explained above,  in the horizon entropy counting we will consider the most general case in which each layer is refined independently, \ie we will set $l=3$. As it will be clear in a moment, considering a subclass of states in which the refinement proceeds in a more synchronized way among the different layers (and which therefore implements also a stronger form of holograpy) simply affects the numerical coefficient in front of the logarithmic corrections, but not the leading linear term.

Taking the large-$n$ limit of \eqref{G}, in order to meet our semi-classicality requirements,  by means of the Stirling formula we get
\be
S_{\mathrm{comb}}^{\max} = \log\left( 
{\mathcal N}\right)=2n l \log{(2)}-\frac{l}{2}\log {(n)}\,.
\ee

It is interesting to note that this is not the first time that Catalan numbers enter the calculation
of black hole entropy, see \cite{Davidson:2011eu}, albeit in a different context.
We will come back to this later.

\section{Full Entropy: from Macro to Micro}

The graph counting quantifies the contribution of the combinatorial degrees of freedom to the 
horizon entropy. This is not the entire entropy of the reduced state.
There is also another, more `geometrical', component which arises from the information 
attached to each vertex by the wavefunction: 
the degeneracy of the Hilbert space for a single vertex, $\Delta(a)$,
for fixed values of the macroscopic quantities. 
This quantity measures the size of the space of wavefunctions compatible 
with our semi-classicality restrictions (and solution of the dynamics equations). 
Since the geometric and combinatorial components are independent, at this (kinematical)  stage
of the construction, the total horizon entropy is then
\be
S(n,a) = \log\left({\mathcal N} \Delta(a)\right)=2nl\log{(2)}+\log(\Delta(a))-\frac{l}{2}\log {(n)}\,.
\ee
This is the entropy of reduced matrix of the most typical state that we can construct
with our special class of condensates, at fixed number of vertices\footnote{If the number
of vertices were not fixed, an additional contribution should be added to include the effect
of the dispersion in the number of vertices.}.

The next step is to maximise this entropy at fixed value of the classical area 
$\aih$ of the horizon. Consider the function:
\begin{equation}\la{ent-funct}
\Sigma(n,a,\lambda) = S(n,a) + \lambda({\aih} - 2 a n)\,,
\end{equation}
where $\lambda$ is a Lagrange multiplier imposing the area constraint. Necessary
conditions for the maximisation of \eqref{ent-funct} are:
\ba
\frac{\partial \Sigma}{\partial \lambda} &=& {\aih} - 2 a n = 0\,, \la{con1}\\
\frac{\partial \Sigma}{\partial n} & \approx& 2l\log (2 ) 
- 2\lambda a =0\,,\la{con2}\\
\frac{\partial \Sigma}{\partial a} & =& \frac{\Delta'(a)}{\Delta(a)}
-  2n\lambda=0\la{con3}\,,
\ea
where the prime indicates derivative w.r.t. $a$. In \eqref{con2} we have dropped a term of 
order $1/n$, since we are in the large $n$ regime, as the semi-classicality conditions of small 
fluctuations of the horizon geometrical properties imply. These are three equations for the 
values of the three quantities $(n,a,\lambda)$ which maximise the entropy functional 
\eqref{ent-funct}. 

We could solve them explicitly only if we knew the expression the 
single vertex Hilbert space degeneracy $\Delta$ as a function of $a$. 
In principle, its evaluation could be achieved by the imposition of the isolated horizon boundary 
condition, by computing the number of solutions to that equation and compatible with the 
dynamics. However, both tasks, \ie solving the isolated horizon boundary condition and the 
equations of motions defining the microscopic dynamics, are highly non-trivial and
currently out of our reach, as we had already emphasised.

Alternatively, we could turn the argument around and use the three relations \eqref{con1},  
\eqref{con2},  \eqref{con3} to get an explicit expression of $a$ and $\Delta(a)$ as functions of $
\aih$  and $\lambda$. 
At this point, then, we could require consistency with a further semi-classical property of a 
black hole horizon, namely its thermality, in order to determine the value of the Lagrange 
multiplier $\lambda$ and thus remove the last ambiguity left in our entropy calculation.
This gives a different twist to this discussion, turning it into the inference of how certain
microscopic quantities should look like if we want them to be compatible with macroscopic
observations.
Following this strategy, we obtain
\ba
&&a=\frac{l\log(2)}{\lambda}\la{a}\\
&&\Delta=c_0\exp{(2\lambda a n)}=c_0\exp{\left(\lambda {\aih}\right)}\la{Delta}\,,
\ea
where $c_0$ is an integration constant left unspecified for the moment.
The horizon entropy we derive is then
\be
S(\aih)\approx 2\lambda {\aih}-\frac{l}{2} \log\left(\frac{\aih}{\ell_{\va P}^2}\right)\,.
\ee
Notice that the solution \eqref{Delta} matches the expectation that the dimension of the Hilbert 
space for the wavefunction at fixed plaquette area should be finite once we implement the 
semi-classicality conditions listed in Section \ref{sec:class}. In particular, this would\begin{enumerate}
\item exclude degenerate (zero-volume) configurations (which would be indeed incompatible with a reasonable semi-classical limit);
\item impose a finite volume of the shell, again consistent with the semi-classical intuition.
\end{enumerate}
In other words, we should expect that whenever $\Delta(a)$ diverges, we should not be 
worried about possible problems in the calculation of the entropy since the geometry would be 
highly problematic to be interpreted as a semiclassical one in the first place.

We can go further in our attempt of determine the properties of states that match the large
scale behaviour of classical gravity. Since the semi-classical limit involves large values 
of $n$, the area constraint \eqref{con1} requires $a$ to be small. In the limit of $a\rightarrow 
0$, the IH boundary condition fixes the holonomy around each radial link to be flat; this can be 
achieved only if the spin labels of the tangent links are 0. Therefore, in the limit  $a\rightarrow 
0$ the wavefunction should be a delta peaked on $j_I=0$, which means $\Delta(0)\sim 1$. As 
soon as $a>0$, then $\Delta (a)$ should grow\footnote{Notice that these arguments about semi-classical limits are used only to constrain the functional dependence of $\Delta(a)$ on $a$; there is no requirement that $j_l$ actually ever take the zero value, which may be problematic from the quantum geometric point of view.}. These expectations are matched by the Taylor 
expansion of the solution \eqref{Delta} around $a=0$, if we fix the integration constant 
$c_0=1$, namely
\be
\Delta(a)\sim 1+\lambda a\,, \la{eq:densityofstates}
\ee
for small $a$.

The free parameter $\lambda$ can now be fixed by requiring consistency with the 
semi-classical thermodynamical condition relating the derivative of the entropy w.r.t. 
a local notion of energy at the horizon for fixed $n$ to the (inverse) Unruh\footnote{A geometric 
notion of  temperature \cite{Pranzetti:2013lma} can be associated to a quantum IH by 
demanding the Kubo--Martin--Schwinger condition \cite{Haag:1992hx} to be satisfied for a 
sub-algebra of the holonomy-flux $*$-algebra of LQG. In the large IH area this notion of 
temperature coincides with the Unruh one.} $1/T_{\va U}=\beta_{\va U}=2\pi /
(\ell_{\va P}^2\kappa) =2\pi \ell/\ell_{\va P}^2$. In the previous expression, $\ell$ is a local 
stationary observer proper distance and $\kappa=1/\ell$ is its surface gravity. For this 
purpose, we can use the local notion of energy introduced in \cite{Frodden:2011eb} for 
isolated horizons, namely
\be
{\mathcal E}_{\va IH}=\frac{\aih}{8\pi\ell}\,.
\ee
Dropping a subdominant term ($\propto 1/\aih$), which becomes immaterial for large enough
areas, we get
\be
\beta_{\va U}=\frac{2\pi \ell}{\ell_{\va P}^2}=\frac{\partial S}{\partial {\mathcal E}_{\va IH}}=8 \pi \ell \frac{\partial S}{\partial \aih}
\approx{16 \pi \ell}  \lambda\,,
\ee
which leads, finally, to
\be
\lambda \approx \frac{1}{8\ell^2_P}\,.
\ee
Therefore, consistency with the IH thermodynamical properties yields an entropy
\be\la{entropy}
S(\aih)\approx  \frac{\aih}{4\ell_{\va P}^2}-\frac{3}{2} \log\left(\frac{\aih}{\ell_{\va P}^2}\right)
\ee
whose leading term reproduces the Bekenstein--Hawking formula; the numerical coefficient in 
front of the sub-leading  logarithmic correction comes from explicitly setting $l=3$ as argued 
above, and it matches the result of previous calculations in LQG 
\cite{Kaul:2000kf, Livine:2005mw, Engle:2011vf} as well as those performed through CFT techniques 
\cite{Carlip:2000nv}.

\section{Conclusions and final remarks}
We now briefly summarise the results obtained and observations made in this paper.

\

We have constructed, within the full quantum gravity formalism of group field theory, a class of quantum states that can be argued to describe (at least some key properties of) continuum spherically symmetric geometries. We have done this by making use of both its quantum geometric aspects, shared with loop quantum gravity, and its combinatorial aspects, shared with random tensor models. We have then imposed additional conditions on such states, that support an interpretation of them as containing horizons, and thus describing black hole geometries. For such states, we have then identified the microscopic degrees of freedom contributing to the horizon entropy and compute the latter explicitly. Under appropriate semi-classicality restrictions and the assumption that the entropy is maximised, we have recovered both a general area law and the exact Bekenstein--Hawking value. The fact that our states are realistic quantum states in the full theory, involving also a highly non-trivial sum over graphs, including arbitrarily refined ones, is one main improvement over the existing derivations of the same results in the loop quantum gravity literature. Let us now mention a few additional interesting aspects of our analysis and results.

\

The entropy result \eqref{entropy} is completely independent on the Barbero--Immirzi 
parameter $\gamma$. This is a striking consequence of the GFT formalism in its Fock 
representation and the way in which it allows us to work with quantum gravity states.
More precisely, the GFT fundamental field operators \eqref{c-field}, encoding 
the state geometrical data in the collective wavefunction $\sigma_r$ for a given shell, 
represents the key departing point from canonical LQG. It allows us to introduce a well-defined 
number operator in the theory and replace area eigenstates and eigenvalues for the quantum 
horizon with condensate states and \emph{area expectation values} on a single vertex state, 
(which is the same for each fundamental block due to the condensate hypothesis). In this way, 
while $\gamma$ still enters in the value of $a$, it completely disappears in the final expression 
of the entropy. Another way to understand the same important point is that this is a consequence of our choice of states in a second quantised formalism (reasonably simple condensates), allowing us to do a microscopic counting of degrees of freedom, while not using eigenstates of the total area operator. The latter,  customary used in LQG, beside being dubious as corresponding to semiclassical black hole states, immediately introduce a dependance on the Barbero--Immirzi  parameter, since this appears inevitably in the area spectrum. This also means that obtaining our results, including this intriguing independence from the Barbero--Immirzi  parameter, have been facilitated greatly by  the GFT formalism, with its convenient organisation of spin network states into a Fock space (with the mentioned existence of a number operator), and the tools it provides for handling sums over different triangulations/graphs.
The reason for this independence from the Barbero--Immirzi  parameter can also be traced back to the appearance of \eqref{Delta} (see the related discussion in the text). It is not just the area value that matters, but also the single vertex `density of 
states'. The calculations that we have given above suggest that this should lead to terms
compensating $\gamma$ in the physical quantities. They also suggest that LQG calculations
might be overlooking an essential step, signalled by the appearance of the Barbero--Immirzi
parameter in expressions for quantities that should be independent of it, at least at the 
classical level. The evaluation of the expectation values of operators on properly defined
semiclassical states (which are not necessarily eigenvectors of the area operator for some
fixed graphs structure) seems to be a critical calculation, in this respect.

\

The graph basis provides, in this case, an orthonormal basis for the reduced density 
matrix. This, together with the holographic property of our states, allows us to compute the 
horizon entanglement entropy exactly and prove the result
\be
S_{Von Neumann}\equiv -\tr{(\rho_{tot, r_0}^{(n)}\log(\rho_{tot, r_0}^{(n)}))}=\log(\mathcal N(n))\equiv S_{\mathrm{Boltzmann}}\,,
\ee
which then reduces to a relatively simple counting exercise. As we have already mentioned,
the particular class of states that we have chosen, obtained by what are very close to be
melonic refinement of seed states, has a very specific growth rate with the number of vertices.
This growth has also been  considered  in \cite{Davidson:2011eu}, in a different context and
with a different picture in mind. Despite all the limitations that have been discussed, these
results point at a specific behaviour for physical quantum gravity states describing black
holes. Together with the observation about the disappearance of the Barbero--Immirzi 
parameter, related to the density of physical states per vertex, these facts suggest that the
superposition of microscopic \emph{kinematical} configurations in the physical states cannot
be neglected in such calculations, and it might also have to be severely constrained.

\

Following up on the last point, the dynamics has not yet been used, essentially.
Our results can be seen as general, therefore, only provided that the states we have used turn out to be good representatives of the true microstates of physical black holes, not only encoding in a more precise manner the various properties that we expect from such particular spacetime geometries (at least semi-classically) and that we have imposed only implicitly, but also solving at least approximately the microscopic quantum dynamics of the theory.

Given the experience in the cosmological setting, one should expect that the imposition of the 
GFT equations of motion will result in some non-linear (set of) effective equation(s) 
on the wavefunction and on the coefficients of the linear
combinations defining our quantum states in terms of graphs.
From the discussion above, it might be relatively easy, once even only a qualitative 
understanding of the structure of some of the solutions to the equations is achieved,
to check the compatibility of the dynamics with macroscopic properties, as quantities
as rate of growth of graphs with the number of quanta are already rather constrained. We leave this type of analysis for future work, though.

The maximum entropy principle that we are used, itself, may or may not be compatible 
with the full quantum dynamics. At this stage, we use it to fix the shape of the states given the available 
information and try to infer consequences about their nature of the states, but nothing more.
It goes without saying that it would be even more significant if we could recover this reasoning
as an endpoint of a calculation that starts from the microscopic equations, especially given
the deep relations between gravity and thermodynamics.

\addcontentsline{toc}{section}{Acknowledgement}
\section*{Acknowledgement}
Research at Perimeter Institute for Theoretical Physics is supported in part by the Government of Canada through NSERC and by the Province of Ontario through MRI.

\appendix
\section{Basics of GFT Fock space}\la{app:GFTFock}
In this Appendix we briefly summarise the necessary ideas about group field theories that
are needed in the main body of the paper, in order to make the text as much self-contained
as possible. We will closely follow \cite{Oriti:2015qva}.
A more thorough discussion can be found in \cite{Oriti:2013aqa,Oriti:2014uga}.

In a nutshell, GFTs are a generalisation of matrix models for two-dimensional quantum gravity.
They are quantum field theories over group manifolds whose perturbative expansion
generates Feynman diagrams whose combinatorial structure can be put in correspondence
with triangulations of higher dimensional manifolds.
Besides extending the combinatorial structure of the Feynman rules, they include additional data
which can be used to store additional geometric information. In particular, these data
allow the construction of models which can be put in correspondence with LQG and
spinfoam models.
While the formalism \emph{per se} leaves considerable freedom for model building, we
will restrict the discussion to the simplest models that are relevant for 3+1 quantum gravity.

One possible presentation of GFTs is the one based on the Fock space representation,
which starts from the idea that the basic building block of a theory, the elementary quantum
excitation, is a single four-valent spin network vertex.
The basic ingredients are group field ladder operators $\hphi, \hphid$, defined over the group
$\SU(2)^4$. 
The argument $g$ of one such field operator is then the tuple 
$(g_{1},g_{2},g_{3},g_{4})$. 

A geometric interpretation is then attached to this type of data: each $\SU(2)$ argument
is interpreted as the parallel transport, induced by a gauge connection, from a four-valent vertex 
to an endpoint of an edge emanating from it. For consistency with this interpretation,
then, a gauge invariance property is enforced in terms of the invariance of the field
operators under diagonal right action of $\SU(2)$:
\begin{equation}
\hphi(g k_{\mathrm{Diag}}) = \hphi(g), \qquad k_{\mathrm{Diag}} = (k, k, k, k),\qquad \forall k\in 
\SU(2)\,.
\end{equation}
Notice that, with a slight abuse of notation, we will refer to this sort of gauge invariance by using
directly  the $\SU(2)$ group elements, even when the diagonal action on $\SU(2)^4$
is intended.

The algebra of \emph{Bosonic} ladder operators is determined then by the commutation relations:
\begin{equation}
\left[
\hphi(g_v), \hphid(g_v')
\right]
= \Delta_R(g_v, g_v')
\end{equation}
where we are using the following definition for the right-invariant Dirac delta distribution on
$\SU(2)^4$:
\begin{equation}
\Delta_R(g_v, g_v') = 
\int_{SU(2)} dh \prod_{I=1}^4 \delta(g_{v,I} h g_{v', I}^{-1})
\end{equation}
and $\delta(g)$ denotes the ordinary delta-function over a single copy of $\SU(2)$.

The role of these ladder operators is then to create and annihilate objects with all the
properties of spin network vertices. Using the non-commutative Fourier transform
on these basic operators gives rise to a representation in which the fundamental
ladder operators correspond to elementary quanta labeled with the data of quantum tetrahedra.

Generic states in the Fock space defined by these representations, then, will correspond
to sets of quantum tetrahedra or spin network vertices variously arranged. Among these
states there will be the ones corresponding to closed connected spin networks, or, equivalently,
triangulated three dimensional manifolds.
The properties of the states can be encoded into wavefunctions, as usual. For instance,
a generic n-particle state can be written as:
\begin{equation}
\ket{\psi} = \int_{(\SU(2)^4)^n} \prod_{i=1}^n \mathrm{d}g_{v_i} 
\left(
\psi(g_{v_1},\ldots,g_{v_n})
\prod_{i = 1}^n
\hphid(g_{v_i})\right) \ket{0}\,,
\end{equation}
where $\ket{0}$ is the Fock vacuum.
The wavefunction $\psi$ provides the bridge between GFT and the language of LQG
\cite{Oriti:2014uga}. In particular, by controlling the dependence on the group elements,
the familiar wavefunctions of closed, gauge invariant spin network attached to graphs
can be recovered.

This kinematic construction is completed by definition of suitable dynamics, which we will
not describe in detail here. Typical choices for the equation of motion obeyed by
a physical state $\ket{\Psi}$ are:
\begin{equation}
\frac{\delta \mathcal{S}}{\delta \overline{\varphi}(g)}[\hphi, \hphid] \ket{\Psi} = 0
\la{eq:gfteom}
\end{equation}
where the action $\mathcal{S}$ is designed in such a way that the Feynman expansion
of the theory gives rise to diagrams that can be put in correspondence with four dimensional
simplicial complexes, decorated with appropriate geometric data. The various kernels
appearing in the field monomials, then, can be adjusted to reproduce spin foam amplitudes.

The advantage of the Fock representation stems from the fact that the equation of motion
\eqref{eq:gfteom} does not generically single out states entirely contained in an eigenspace
of the Fock number operator. The Fock representation of the theory makes it very easy
to work with states which are superpositions of states with different number of quanta,
leveraging on the techniques familiar in the domains of Quantum Optics and Condensed
Matter Physics.
The ability of doing so is crucial if we want to obtain even an approximate description of
the continuum regime following from \eqref{eq:gfteom}, where the notion of elementary
quanta dissolves in the continuum soup.

\bibliographystyle{apsrev-title}


\end{document}